\documentclass[final,5p,times]{elsarticle} 

\usepackage[T1]{fontenc}
\usepackage[utf8]{inputenc}
\usepackage{lmodern}
\usepackage{graphicx}
\usepackage{amsmath}
\usepackage{ulem}
\usepackage{cuted}

\def\HgCdTe{\mbox{Hg$_{1-x}$Cd$_x$Te}}
\def\PbSnTe{\mbox{Pb$_{1-x}$Sn$_x$Te}}
\def\Im{\mbox{Im}}
\def\Re{\mbox{Re}}

\def\eff{\mbox{\scriptsize eff}}

\begin{document}

\begin{frontmatter}

\title{Explicit Fresnel formulae for the absorbing double-negative metamaterials}

\author{Igor Tralle}
\address{College of Natural Sciences, Institute of Physics, University of Rzesz\'ow, Pigonia 1, 35-310 Rzesz\'ow, Poland}

\author{Levan Chotorlishvili}
\address{Institut f{\"u}r Physik, Martin-Luther Universit{\"a}t Halle-Wittenberg, D-06120 Halle/Saale, Germany}

\author{Pawe\l\ Zi\c{e}ba}\ead{Corresponding author: I. Tralle, email:pawel.zieba@energybis.pl}
\address{Energy Business Intelligence Systems (energyBIS), Pi\l{}sudskiego 32, 35-001 Rzesz\'ow, Poland}

\begin{abstract}
 We inspect the	optical properties of dissipative double-negative metamaterials (DNMM)  and find explicit expression for the total reflection angle and the correct Fresnel formulae describing the reflection and refraction for the DNMM at the oblique incident of the electromagnetic wave on the interface for \textit{TE} as well as \textit{TM} electromagnetic  wave polarization. The reflectivity and transmissivity of DNMM film embedded in a  positive refraction index (PIM) 
surrounding  are  presented and discussed.
\end{abstract}


\date{\today}

%\maketitle

\end{frontmatter}

\section{Introduction}

In recent years, we have been witnessed of the explosion of interest in a field of research, which is termed metamaterials. This area of research is characterized by an exponential growth of a number of publications, to mention just a few, there are two monographs \cite{cai2010, solymar2009} and the references therein. According to \cite{cai2010}, the term “metamaterials” can be used in a more general, as well as in a more specific sense. In the more general sense, these are materials possessing “properties unlike any naturally occurring substance” or simply “not observed in nature.” More specifically, these are the materials with a negative refractive index, whose existence and properties were discussed for the first time by Veselago \cite{veselago1968}.

It is worth mentioning that most of the proposed ever since designs of metamaterials were characterized by ever increasing sophistication of fabrication methods. Contrary to these, in our previous publications \cite{tralle2014,pasko2017}, we proposed a relatively simple way to  fabricate a three-component artificial composite metamaterial and demonstrated by numerical simulations, what are the domains of its existence. It means, that  we set seven independent parameters such as temperature, external magnetic field, relative concentration of ingredients and some others to be controlled at the numerical simulations. In other words, we have seven-dimensional parameter space to search through, in order to establish the frequency domain where this composite material becomes metamaterial with negative refractive index.

For the readers' convenience, here we outline briefly the main ideas. Let us assume  we have a mixture of three materials, and each has granular or powder form, such that the grain sizes are much smaller than the electromagnetic wavelength propagating in the medium. We wish to match the properties of ingredients in such a way that the effective dielectric permittivity of the composite would be determined by the three components, while its magnetic permeability only by one of them, responsible for the magnetic properties of the mixture. This third ingredient by the assumption should determine the effective permeability of the hypothetical material. Suppose it to be metal magnetic nano-particles (or grains; we shall use these two words interchangeably). We treat these metallic grains as immersed or dispersed in a weakly conducting matrix. If the metallic particles are supposed to be single-domain, then we can take into account only the orientation alignment of their intrinsic magnetic moments and do not need to take into account their induced magnetic moments, as it can be proved (see \cite{landau1960}, Chap. 82). The sizes of the single-domain particles depend on the material and contributions from different anisotropy energy terms. If we assume nano-particle shape to be spherical, then typical values for the critical radius $a$ are about 15 nm for Fe and 35 nm for Co, for $g$ – Fe$_{2}$O$_{3}$ it is about 30 nm, while for SmCo$_{5}$, it is as large as 750 nm \cite{givord1991}. Now we can treat the suspension of metallic grains as a kind of "frozen paramagnetic macromolecules," where the metallic nano-particles play the role of "macromolecules."

The magnetic moments of these single-domain nano-particles at room temperature are randomly distributed and we can describe their behavior in the framework of Langevin theory of paramagnetism. Note that the 'swarm' of magnetic nanoparticles immersed into another medium was already considered in scientific literature and even the term for describing this situation was already coined, namely, superparamagnetism. The point is that such system behaves like a paramagnet, with one notable exception that the independent moments are not that of a single atom, but rather of a single-domain ferromagnetic particle, which may contain more than $10^5$ atoms. In the absence of an external magnetic field their magnetic moments are distributed at random,  but  being placed in magnetic field, magnetic moments of individual grains treated in terms of classical physics start to precess, that is why the frequency range in which  $Re [\mu_{\mbox{eff}}(\omega)]$ is negative, appears in the vicinity of resonance $\omega_0 \approx \omega$, where $\omega$ is the frequency of the electromagnetic wave incident of the medium and $\omega_0 = \gamma H_0$. Here $\gamma$ - is the gyromagnetic ratio and $H_0$ -external magnetic field.

It turns out, for the composite to become a  metamaterial it is important that the sizes of ferromagnetic nanoparticles, their magnetic moments, the relative concentration of the ingredients, Cd- or Sn-content in the semiconductor compounds, the temperature, and the external magnetic field have to have certain definite values. In our previous works it was  shown that the mixture composed of a 'swarm' of single-domain ferromagnetic nanoparticles, small metallic particles (Ag, or Al or Cu) and the small semiconductor particles of \HgCdTe, or \PbSnTe) attains double-negative metamaterial properties in  the frequency range  $10-100$ GHz being placed in an external magnetic field. The reasoning behind this choice of Hg$_{1-x}$Cd$_{x}$Te, or Pb$_{1-x}$Sn$_{x}$Te is the following. The electrical properties of these materials crucially depend on cadmium (the same is about Sn, but for definiteness let speak of  Hg$_{1-x}$Cd$_{x}$Te and  concentration of cadmium $x$. If $x = 0$, that is in case of HgTe, the material is semimetal with energy gap $E_g < 0$, while in case of  $x = 1$ (CdTe) material becomes semiconductor with wide energy gap of about 1.5 eV at 300 K. Thus, changing the concentration of cadmium, one can change the energy gap, and hence the concentration of free electrons. In terms of our model, it means that one can pass smoothly and continuously from Lorentz model for dielectric permittivity, where the electrons are almost tightly bounded to Drude model, where they are almost free to move. As a result, cadmium concentration becomes an important parameter of the model; by means of it— among others— one can control the frequency range where the real part of dielectric permittivity can be made negative and force it to overlap with the frequency domain, where magnetic permeability is negative.

Here we use the term 'double-negative' to emphasize the negativity of the real parts of permittivity and permeability,that implies the composite refractive index $\Re[\tilde{n}]$ is negative \cite{veselago1968}. The negativity of the real part of refraction index  entails that the permittivity and the permeability are complex-valued functions  \cite{ziolkowski2001,milonni2005}, meaning the refraction index is also complex-valued.

Metamaterials may be engineered to exhibit a negative refraction  \cite{pendry1999, shelby2001, engleta2006, solymar2009, cai2010} but they tend to be absorptive and narrow-band for the fundamental reasons \cite{stockman2007,kinsler2008}, albeit  the imaginary part can be relatively small as long as it is allowed by the causality condition.

Despite of this, many authors who explore metamaterials  often treat them as non-absorbing (see, for instance, \cite{veselago1968, ceji2005, veselago2003}). To have non-absorbing metamaterials is very desirable since they promise numerous possibly very interesting and exciting applications \cite{kildishev2011}. It is however \textit{logically inconsistent}  to treat them as non-absorbing for the very simple reason: as it was mentioned above, one can assign minus sign to the refraction index  only if permittivity and permeability are \textit{complex-valued} functions. In fact, the term \textit{negative  refraction index} is used for short and one should keep in mind that it in fact, \textit{negative real part of refraction index} is talked about, and that its imaginary part is positive and metamaterial is always absorbing.

In the classical optics of non-absorbing media one often deals with plane periodic electromagnetic waves whose planes of constant phase and amplitude are normal to the wave vector. Such waves were designated by Voight as homogeneous waves. In absorbing media another type of waves appears for which the planes of constant phase and amplitude are no longer parallel; Voight designated them as inhomogeneous waves. 
It is interesting to note that a relatively small number of papers were devoted to the treatment the reflection and refraction of electromagnetic wave at oblique incidence on the interface between non-absorbing and absorbing media, among them the papers by \cite{mahan1956, dupertuis1994}. It is also worth mentioning the paper devoted to oblique surface waves at an interface \cite{slobodan2012}.

There are many papers and textbooks devoted to the propagation of electromagnetic waves in metamaterails (see for example, \cite{ziolkowski2001}, and the books mentioned above\cite{engleta2006, solymar2009}), but there is very little information if ever, concerning the reflection and refraction of electromagnetic wave at oblique incidence on the interface between positive refraction index material (PIM) and  metamaterial. So, the  main goal of present work is to derive the explicit  Fresnel formulae for the gyrotropic, magnetic, birefringent and absorbing metamaterials  and  the study of  optical properties of such material at the oblique incidence of electromagnetic wave on the interface  between them and the positive refraction index material. 

At the end of  Introduction we would like to add some comments concerning terminology we use throughout the paper. Till now there is no unanimity as for this subject is concerned: some authors use the term ’negative group velocity materials’, some others  prefer the term ’negative phase velocity materials’. This is because the phase velocity and group velocity are directed against each other in case of such materials, and which direction is positive and which is negative is a matter of convention.  For that reason we use the term double-negative metamaterial (DNMM) throughout, in order to emphasize that in this case the real parts of dielectric permittivity as well as magnetic permeability are simultaneously negative in some frequency domain.
This paper is organized as follows.  In  section \textbf{II} we  discuss the magnetic birefringence and Faraday effect in DNMM, in  section \textbf{III} we treat the wave propagation in DNMM and in  section \textbf{IV} we derive Fresnel formulae for it while in \textbf{V} we consider  the reflectivity and transmittance of the DNMM-films.

\section{Magnetic birefringence and Faraday effect in DNMM}

Metamaterial proposed in \cite{tralle2014,pasko2017} is interesting from several points of view. First, it is not a complicated engineering construction, but the mixture of three ingredients. Second, despite the fact that it is not a crystal, it is anisotropic, optically bi-axial medium displaying Faraday effect. Third, its effective dielectric permittivity (we call it effective, because it is the permittivity of mixture) can be considered as a complex scalar, while effective magnetic permeability relates to the permeability of the third ingredient of the mixture, that is the 'swarm' of ferromagnetic nano-particles via  
the tensor represented by 3-by-3 non-Hermitian matrix. As it was shown in \cite{tralle2014}, this matrix is of the form:  
\begin{displaymath} \tilde{\mu} = 
\left( \begin{array}{ccc}
1+ 4\pi\chi &  4\pi(iG)  & 0 \\
-4\pi(iG)  & 1 + 4\pi\chi & 0 \\
0    &  0   & 1 
\end{array}\right),
\end{displaymath}
where 
\begin{displaymath}
\chi = \chi_0 \frac{\omega_0^2}{2i\Gamma} \left(\frac{1}{\tilde{\omega_1}-\omega} - \frac{1}{\tilde{\omega_2} + \omega }\right),\hspace{1cm}\\  
\end{displaymath}

\begin{displaymath}
G = \chi_0 \frac{\gamma\omega}{2i\Gamma} \left(\frac{1}{\tilde{\omega_1}-\omega} - \frac{1}{\tilde{\omega_2} + \omega }\right)H_0 .
\end{displaymath}
 Here $\Gamma = \tau^{-1}$, $\tilde{\omega}_1 = -i\Gamma + 
\sqrt{\omega_0^2 - 2\Gamma^2}, \tilde{\omega}_2 = -i\Gamma - \sqrt{\omega_0^2 - 2\Gamma^2}$ and $\omega_0 =\gamma H_0$, where $\tau$ is the magnetic moment relaxation time (see \cite{tralle2014} for details) and $\bf{H}_0$ is external magnetic field. \\

If the wave vector of incident electromagnetic wave aligned arbitrary with respect to external magnetic field, effective permeability is a tensor and the medium is a\-ni\-so\-tro\-pic. However, if we consider the simplest case when $\mathbf{k}\parallel\mathbf{H}_0$,  then two circularly polarized waves can propagate in such medium for which the magnetic permeability and hence, the refraction indices  are different (one for 
left-polarized wave, while the other for the right-polarized one). If one introduces the following auxiliary quantities
$\chi_{\pm} = \chi \pm G $ , then the refraction indices for these two waves are  $\tilde{n}_{\pm} = \sqrt{\tilde{\epsilon}_{\eff}\tilde{\mu}_{\eff,\pm}}$, 
where $\tilde{\mu}_{\eff,\pm}$  and $\tilde{\epsilon}_{\eff}$ stand for the effective magnetic permeability and effective dielectric permittivity of metamaterial, respectively. The effective  dielectric permittivity and magnetic permeability were calculated in the framework of Bruggeman approximation often called  the effective medium theory \cite{bruggeman1935}. Its main asset is that all ingredients of a mixture by assumption are treated on the same footing in a symmetric way and none of them plays a privileged role. For example, the effective dielectric permittivity $\tilde{\epsilon}_{\eff}$ is calculated as the root of the following third-order algebraic equation:
\begin{eqnarray}
&&f_1 \frac{\epsilon_1(\omega) - \tilde{\epsilon}_{\eff}(\omega)}{\epsilon_1(\omega) + 2\tilde{\epsilon}_{\eff}(\omega)} + \\
&& + f_2 \frac{\epsilon_2(\omega) - \tilde{\epsilon}_{\eff}(\omega)}{\epsilon_2(\omega) + 2\tilde{\epsilon}_{\eff}(\omega)} + f_3 \frac{\epsilon_3(\omega) - \tilde{\epsilon}_{\eff}(\omega)}{\epsilon_3(\omega) + 2\tilde{\epsilon}_{\eff}(\omega)} =  0, \nonumber
\end{eqnarray}
where $\epsilon_i, i=1,2,3$ are the dielectric permittivities of three ingredients of the mixture $f_i$ is the volume filling fraction of the $i-th$ material in the mixture. Obviously, these three quantities in a natural way obey the following additional condition:$f_1 + f_2 + f_3 = 1$. We calculated the roots of the equation numerically, because they depend on the concentrations of the constituent components of mixture. They were not known beforehand and it was more convenient from computational point of view to solve this equation numerically.  Since we consider absorbing medium, we always chosen the root that had positive imaginary part. 
Since the magnetic permeabilities of two components other than 'swarm' of ferromagnetic nanoparticles, are equal to $1$ in wide frequency range, the expression for effective magnetic permeability of the mixture takes more simple form, namely $\mu_{\mbox{eff},\pm} = f_{12} +(1 - f_{12}(1 +  4\pi\chi_{\pm}(\omega))$, where $f_{12} = f_1  +  f_2$.

It is interesting and worth noting that in this case real part of refraction index is negative only for one of  two waves propagating in a medium. The case of arbitrary $\bf{k}$-vector alignment  with respect to $\bf{H}_0$, is more complicated and will be considered elsewhere. here we simply state that propagation of the electromagnetic wave proceeds with two different phase velocities  $v_+ = c/ |\Re[\tilde{n}_{+}]|$  and  $v_- = c/ |\Re[\tilde{n}_{-}]|$, where $c$ is light velocity in vacuum.
Assuming  $\Re[\tilde{n}_+] > \Re[\tilde{n}_-]$ we conclude that it takes more time for “slower” wave to traverse the plate made of our material.  The time delay between the two waves while traversing the plate or slab made of such material, is equal to ($d$ is the slab thickness)
	\begin{equation}
\label{time delay}
\Delta t =
	d\left(
	\frac{1}{v_+} - \frac{1}{v_-}
	\right) =
	\frac{d}{c}
		\left(
			|\Re[\tilde{n}_{+}]| - |\Re[\tilde{n}_{-}]|\right).
	\end{equation}
 
For both waves the total revolution of $\mathbf{E}$ or $\mathbf{H}$ vector lasts for the wave period $T = 2\pi/\omega$, meaning that the retarded wave (assuming $|\Re[\tilde{n}_+]| > |\Re[\tilde{n}_-]|$) arrives at the opposite surface of the slab end with the $\mathbf{E}_-$ vector revolved at a larger angle than the other $\mathbf{E}_+$. The difference in rotation angle is 

	\begin{equation}
\label{difference in rotation}
\Delta\alpha = 2\pi\frac{\Delta t}{T} = \frac{2\pi d}{\lambda_0}\left(|\Re[\tilde{n}_{+}]| - |\Re[\tilde{n}_{-}]|\right),
	\end{equation}
where $\lambda_0 = cT$. Note that the rotation of the polarization plane in this case is linearly proportional to the magnetic field parallel to the direction of wave propagation and hence, it is nothing else but the Faraday effect.  
Having in mind that  
$\tilde{n}_{\pm}=$ $\sqrt{\tilde{\epsilon}_{\eff}\tilde{\mu}_{\eff,\pm}}$
and
$\tilde{\epsilon}_{\eff} = \epsilon^{'}_{\eff} + i\epsilon^{''}_{\eff}$,
$\tilde{\mu}_{\eff,\pm} = \mu^{'}_{\eff,\pm} + i\mu^{''}_{\eff,\pm}$,
where $\tilde{\epsilon}_{\eff}, \tilde{\mu}_{\eff,\pm}$
stand for the effective permittivity and permeability of the material we are talking about,
we drop henceforth the subscript $\mathrm{\eff}$ and find the  expressions

\begin{gather}
\label{refraction index}
\tilde{n}_{\pm} =
	\sqrt{|\tilde{n}^2_{\pm}}|\exp( i\phi_{\tilde{n}}),\\
\sqrt{|\tilde{n^2}_{\pm}}| =
	\left[\left(\epsilon^{'}\mu^{'}_{\pm}\right)^2 + \left(\epsilon^{''}\mu^{''}_{\pm}\right)^2\right]^{1/4},
\end{gather}
and
\begin{equation}
\label{refraction angle}
\phi_{\tilde{n}} =
\frac{1}{2}\arctan\left(\frac{\mu^{'}_{\pm}\epsilon^{''}+\epsilon^{'}\mu^{''}_{\pm}}{\mu^{'}_{\pm}\epsilon^{'} - \epsilon^{''}\mu^{''}_{\pm}}\right) \in (\pi/2, \pi),
\end{equation}
from which the explicit expressions for $\Re[\tilde{n}_+]$  and $\Re[\tilde{n}_-]$ follow. Since the functions $\chi(\omega), G(\omega)$ are complex-valued, the absorption of left-polarized and right-polarized waves are a bit different; this different absorption of the right and left circularly polarized light is known as magnetic-cicular dichroism. As a result, the initially linearly polarized wave which is the superposition of left- and right-polarized waves acquires during its propagation within such medium some ellipticity that is, becomes elliptically polarized.

\section{Total reflection angle in case of absorbing DNMM}

Now we consider the process of wave propagation in such material starting from Maxwell equations, in order to study the reflection and refraction on the boundary between two media, the first one is of positive refraction index (PIM) and the other one is double negative
metamaterial (DNMM). It should be noted that the optical properties of absorbing materials were considered already by many authors, for example by Born and Wolf in their classical book \cite{born2005} and using somewhat different approach, by M.A. Dupertuis, M. Proctor and B. Acklin \cite{dupertuis1994}, as well as quite recently by P.C.Y. Chang, J.G. Walker and K.I. Hopcraft \cite{chang2005}. All these authors (Born and Wolf including) considered however absorbing, PIM and nonmagnetic materials and hence, they assumed $\mu = 1$. On the other hand, the authors who considered DNMM, treated them as non-absorbing (for instance \cite{ceji2005}).\\

For studying the wave propagation,   reflection and refraction at the boundaries between  two media, one  of  positive refraction index (PIM) and another one, double negative metamaterial (DNMM) one has to inspect the corresponding solutions to the 
the Maxwell equations. As we mentioned in the previous section, in case of considered metamaterial when the wave vector of incident electromagnetic wave aligned arbitrary with respect to external magnetic field, permeability is a tensor and the medium is anisotropic. However, if we consider the simplest case when $\mathbf{k}\parallel\mathbf{H}_0$, one can consider medium as if it would be isotropic with two different values of refractive index for two waves. Keeping this in mind, for a homogeneous, absorbing,  isotropic, linear, charge-free magnetic medium one can write down the Maxwell equations as:
\begin{eqnarray}
\label{Maxwell equations}
  \nabla \times \mathbf{E}(r,t) &=& -\tilde{\mu}\partial_t \mathbf{H}(\mathbf{r},t), \nonumber\\
  \nabla\cdot\mathbf{E}(\mathbf{r},t) &=& 0,  \\
  \nabla \times \mathbf{H}(r,t) &=&\epsilon\partial_t \mathbf{E}(\mathbf{r},t) + \sigma\mathbf{E}(\mathbf{r}, t), \nonumber\\
 \nabla\cdot\mathbf{H}(\mathbf{r}, t) &=& 0. \nonumber
\end{eqnarray}
 Here $\tilde{\mu}$ is the complex permeability, $\epsilon$ is the real permittivity and $\sigma$ stands for the conductivity of the medium. 
 For absorbing media such as metals the  wave vector is complex \cite{born2005}: $\mathbf{k}=k_1\mathbf{e_1} + i k_2\mathbf{e_2}$. Searching for  a  plane wave solution 
\begin{eqnarray}
\label{wave equation1}
&& \mathbf{E}(\mathbf{r}, t)=\mathbf{E_0}\exp(i\mathbf{k}\cdot\mathbf{r} - i\omega t), \nonumber\\
&& \mathbf{H}(\mathbf{r}, t)=\mathbf{H_0}\exp(i\mathbf{k}\cdot\mathbf{r} -i\omega t),
\end{eqnarray}
one gets: 
\begin{eqnarray}
\label{wave equation2}
&&  \mathbf{k}\times\mathbf{E_0} = \tilde{\mu}\omega\mathbf{H_0},\hspace{1em} \mathbf{k}\cdot\mathbf{E_0} = 0, \nonumber\\
&&  \mathbf{k}\times\mathbf{H_0} = -\tilde{\epsilon}\omega\mathbf{E_0},\hspace{1em} \mathbf{k}\cdot\mathbf{H_0} = 0,
\end{eqnarray}
where $\tilde{\mu}=\mu^{'} + i\mu^{''}, \tilde{\epsilon} = \epsilon + i\sigma/\omega = \epsilon^{'} + i \epsilon^{''}$ and $\mathbf{k}$ 
are now complex numbers. Remembering all the time that material which we consider is birefringent, we nevertheless dropped the subscripts $\pm$ in what follows in order to make formulae more readable. At the end of calculations one can simply choose the corresponding subscript $+$ or $-$.
Complex vectors, like  in Eq.~(8) sometimes are called  bi-vectors.
Then, from Eq.~(\ref{Maxwell equations}) one can infer
\begin{eqnarray}
\label{wave equation3}
(\mathbf{k}\cdot\mathbf{k})\mathbf{E} = -\tilde{\mu}\tilde{\epsilon}\omega^2\mathbf{E},
\end{eqnarray}
and thus
\begin{equation}
\label{wave equation4}
k^2 = k^2_1 - k^2_2 + 2i\mathbf{k_1}\cdot\mathbf{k_2} + 2i\cos(\mathbf{e_1}, \mathbf{e_2}).
\end{equation}
The $EM$-waves with this property are called non-uniform (or inhomogeneous) $EM$-waves and of course, they were already considered in the literature \cite{born2005,pincherle1947}. The equations  $\mathbf{k_1}\cdot\mathbf{r} = \mbox{const}$ and  $\mathbf{k_2}\cdot\mathbf{r} = \mbox{const}$    determine the planes of equal phases and equal amplitudes, respectively.  It is convenient to introduce in what follows  the relative and dimensionless complex permittivity and permeability by
$
\tilde{\mu}\tilde{\epsilon}\omega^2 =
	\epsilon_0 \mu_0 \tilde{\epsilon}_r\tilde{\mu}_r\omega^2 = k^2_0\tilde{\epsilon}_r\tilde{\mu}_r,
$
where $\epsilon_0$ and $\mu_0$ are the permittivity and permeability of vacuum, while the subscript $r$ is for 'relative'.
If losses are negligible one can define the wave phase velocity in the medium as $v_ph = c/n =\omega/k$ and the wave number as $k=(\omega/c)n$.
For  absorbing media it follows: 
$
\tilde{k}=(\omega/c)\tilde{n}$. Since $\tilde{n}^2 = \tilde{\epsilon}_r\tilde{\mu}_r,
$
from these and Eq.(\ref{wave equation4}) one deduces
\begin{eqnarray}
\label{wave equation5}
&& k^2_1 - k^2_2 = k^2_0\left[\left(n^{'}\right)^2 - \left(n^{''}\right)^2\right],\nonumber\\
&& \mathbf{k_1}\cdot\mathbf{k_2} = k_1 k_2 \cos(\mathbf{e_1}, \mathbf{e_2})= k^2_0\left(n^{'}\right)\left(n^{''}\right),
\end{eqnarray}
and
\begin{eqnarray}
\label{wave equation6}
&& \left(n^{'}\right)^2 - \left(n^{''}\right)^2 =\left(\epsilon^{'}_r\mu^{'}_r - \epsilon^{''}_r\mu^{''}_r\right), \nonumber\\
&& \left(n^{'}\right)\left(n^{''}\right)= (1/2)\left(\epsilon^{'}_r\mu^{''}_r + \epsilon^{''}_r\mu^{'}_r\right).
\end{eqnarray}
Equations (\ref{wave equation6}) imply 
\begin{eqnarray}
\label{reflection and refraction}
\left(n^{'}\right)^2 &= &
\frac{
	\left(
		\epsilon^{'}_r\mu^{'}_r
	-
		\epsilon^{''}_r\mu^{''}_r
	\right)^2}{2} \nonumber\\
&&+ \frac{
	\sqrt{
			\left(
				\epsilon^{'}_r\mu^{'}_r - \epsilon^{''}_r \mu^{''}_r
			\right)^2
			+ \left(
				\epsilon^{'}_r\mu^{''}_r + \epsilon^{''}_r\mu^{'}_r
			\right)^2
		}
	}{2}, \nonumber\\
n^{''}  &= &
		\frac{
			\left(\epsilon^{'}_r\mu^{''}_r + \epsilon^{''}_r\mu^{'}_r\right)
		}{2n^{'}}.
\end{eqnarray}
The next interesting issue is the reflection and refraction of $EM$-wave on the boundary between positive refraction index material (PIM) and double-negative metamaterial (DNMM). In \cite{veselago2003} the Snell’s law was assumed to be  valid for the non-absorbing metamaterials.
By analogy with non-absorbing dielectric one can write the law of refraction as follows
\begin{eqnarray}
\label{refraction angle2}
\sin\theta_t = \frac{1}{\tilde{n}}\sin\theta_i,
\end{eqnarray}
where the subscripts $i$ and $t$ correspond to the incident and the refracted waves respectively. Due to the complex refraction index, $\theta_t$ is also complex and
cannot be interpreted simply as a refraction angle.  In order to use the complex refraction index, one may resort to ansatz elaborated for the absorbing materials \cite{born2005}. However, one should remember that the material we are dealing with is magnetic and contrary to the case of Born and Wolf \cite{born2005}, permeability in our case $\tilde{\mu}_{\eff,\pm}\neq 1$ but is a complex-valued function.

Let the plane of incidence be the $x-z$ plane.
Then the space-dependent part of the wave phase in our absorbing material
is equal to $\tilde{k}\mathbf{r}\cdot\mathbf{e}^t$.
Here the superscript $t$ stands for 'transiting', that is refracted wave, while the subscripts $x$ and $z$ denote the corresponding components of $\mathbf{e}^t$, the unit vector in the direction of the transmitted wave. Then,

\begin{eqnarray}
\label{Born and Wolf}
&& e^t_x = \sin\theta_t = \frac{n^{'} - i n^{''}}{\left(n^{'}\right)^2 + \left(n^{''}\right)^2}\sin\theta_i, \nonumber\\
&& e^t_z = \sqrt{1 - \sin^2\theta_t},
\end{eqnarray}
and we infer
%\begin{widetext}
\begin{gather}
e^t_x = \frac{1- i\delta}{\Re\left[\tilde{n}\right]\left(1 + \delta^2\right)}\sin\theta_i,\nonumber\\  
e^t_z = \sqrt{1 - \frac{\left(1-\delta^2\right)\sin^2\theta_i}{\left(\Re\left[\tilde{n}\right]\right)^2\left(1 + \delta^2\right)^2} +i\frac{2\Im\left[\tilde{n}\right]\sin^2\theta_i}{\left(\Re\left[\tilde{n}\right]\right)^3\left(1 + \delta^2\right)^2}}.
\end{gather}
\label{widetext}
%\end{widetext}
 $\delta = \frac{\Im\left[\tilde{n}\right]}{\Re\left[\tilde{n}\right]}$
is the "figure of merit".

As in \cite{born2005}, we express $e^t_z$  as $e^t_z = \cos\theta_t= q\exp(i\gamma)$, where

\begin{eqnarray}
\label{generalization of Born and Wolf1}
q^2\cos2\gamma =
	\frac{
		\left(1-\delta^2\right)\sin^2\theta_i
	}{
		\left(\Re\left[\tilde{n}\right]\right)^2\left(1 + \delta^2\right)^2},
\end{eqnarray}
\begin{eqnarray}
\label{generalization of Born and Wolf2}
q^2\sin2\gamma =
	\frac{
		2\Im\left[\tilde{n}\right]\sin^2\theta_i
	}{
		\left(\Re\left[\tilde{n}\right]\right)^3 \left(1 + \delta^2\right)^2
	}.
\end{eqnarray}
From Eqs.~(\ref{generalization of Born and Wolf1}-\ref{generalization of Born and Wolf2}) it follows:
\begin{gather}
	\tilde{k}\left(\mathbf{r}\cdot\mathbf{e_{\tilde{k}}}\right)  =
	\frac{\omega}{c}
	[
		x\sin\theta_i + z\Re\left[\tilde{n}\right]q\left(\cos\gamma - \delta\sin\gamma\right)
    \\
		+ izRe\left[\tilde{n}\right]q\left(\delta\cos\gamma + \sin\gamma\right)
	],
\end{gather}
and for $q^2$ and $\gamma$  
\begin{gather}
\label{generalization of Born and Wolf3}
q^2 = \frac{2\Im\left[\tilde{n}\right]\sin^2\theta_i}{\Delta\sin2\gamma}, \\
	\gamma=\frac{1}{2}\arctan\left(\frac{2\Im\left[\tilde{n}\right]\sin^2\theta_i}{\Delta - \left(1-\delta^2 \right)\sin^2\theta_i}\right),
\end{gather}
\begin{eqnarray}
\label{generalization of Born and Wolf4}
	\Delta = \left(\Re\left[\tilde{n}\right]\right)^2\left( 1+\delta^2\right)^2.
\end{eqnarray}

The obtained Eqs.~(\ref{generalization of Born and Wolf1}-\ref{generalization of Born and Wolf4})  generalize the
 classical formulae by Born and Wolf  for metamaterials.
The constant amplitude planes are defined by the relation $z = \mbox{const}$, meaning  they are parallel to the boundaries  between the two media.
The planes of constant real phase are determined  by the equation
\begin{eqnarray}
\label{The planes1}
x\sin\theta_i + z\Re\left[\tilde{n}\right]q\left(\cos\gamma - \delta\sin\gamma\right)=\mbox{const}.
\end{eqnarray}
These are planes with  normals making  an angle $\theta^{'}_t$  with the normal to the boundary plane
\begin{eqnarray}
\label{The planes normals}
\cos\theta^{'}_t =
	\frac{
		\Re\left[\tilde{n}\right]
		q
		\left(
			\cos\gamma - \delta\sin\gamma
		\right)
	}{
		\sqrt{
			\sin^2\theta_i +
				\left(
					\Re | \left[\tilde{n}\right]
				\right)^2
				q^2(\cos\gamma - \delta\sin\gamma)^2
		}
	}.
\end{eqnarray}
Since the amplitude plane is parallel to the boundary, the angle between the vectors $\mathbf{e_1}$ and $\mathbf{e_2}$ (cf. Eq.(\ref{wave equation4}) ) is equal to  $\theta^{'}_t$.\\

As for the possible applications of metamaterials, it is worth mentioning that there are some papers published already, in which the authors proposed to use these materials to construct the waveguides  \cite{atakara2012, atakara2013, de-kui2004}. The authors of these papers  considered the propagation of TE and TM modes in the waveguide made of DNMM, but for the sake of simplicity, they consider the MTM to be lossless, with the permittivity and permeability tensors taking only real values, which is not very realistic.  As it is known, the operation principle of waveguides is the EM-wave total internal reflection. That is why, it seems  useful to derive  formula for the angle of total internal reflection in case of absorbing metamaterial.  Usually, the discussion of this issue concerns mainly  the region of the Goos-H{\"a}nchen shift at the boundary between PIM and magnetic DNMM \cite{lakhtakia2003}, but not  the total internal reflection angle. The cause of total reflection in metamaterials is the same as in case of usual dielectrics, but in case of metamaterials the formula describing critical angle is more complicated, as it is shown below. \\

Suppose that the EM wave impinges  on the boundary between PIM and DNMM at the side of DNMM. Assume also that DNMM is more optically dense, that is
$\Re \left[	\tilde{n}_{DNMM}\right] > n_{PIM}.$
Writing  the refraction law in the form
$\mathbf{n}\times\mathbf{k}_i=\mathbf{n}\times\mathbf{k}_t$
where $\mathbf{n}$  is the vector normal to a boundary and noting that
$|\mathbf{k_i}| = \tilde{k}_{DNMM}$, $|\mathbf{k_t}|=k_{PIM}$  and
\begin{eqnarray}
\label{The planes2}
&& k_{1,DNMM}\sin\alpha^{'}_i + ik_{2,DNMM}\sin\alpha^{''}_i =\nonumber\\
&& k_{PIM}\sin\alpha_t,
\end{eqnarray}
we  infer the  expression for the critical angle $\alpha^{'}_{i,c}$
(subscript \textit{c}  stands for 'critical'):
\begin{equation}
\sin\alpha^{'}_{i,c} = \frac{k_{PIM}}{k_{1,DNMM}}.
\end{equation}
With Eq.(\ref{wave equation5})-(\ref{reflection and refraction}) and  Eq.(\ref{The planes normals}) one can  calculate the angle of total internal reflection for our case.
Using Eq.(\ref{wave equation5}) we deduce
\begin{eqnarray}
\label{The planes3}
k_1 =\left[\frac{b+\sqrt{b^2 + 4\beta^4 r^2}}{2\beta^2}\right]^{1/2},\hspace{1em}
k_2 =\frac{k^2_0\left(n^{'}\right)\left(n^{''}\right)}{k_1\beta},
\end{eqnarray}
where $\beta = \cos\left(\mathbf{e_1}, \mathbf{e_2}\right)$,
$b = k^2_0\left[\left(n^{'}\right)^2 - \left(n^{''}\right)\right]\beta^2$,
$r^2 = \left[k_0^2\left(n^{'}\right)\left(n^{''}\right)\right]^2 $ and $n^{'}, n^{''}$ are determined by means of Eq.(\ref{reflection and refraction}).
Keeping in mind that Eq.(\ref{The planes2}) $k_{1,DNMM}$ is simply equal to $k_1$ from  Eq.(\ref{The planes3}) and $\beta =\cos\left(\mathbf{e_1}, \mathbf{e_2}\right)=\cos\theta^{'}_t$, one can determine the angle of total internal reflection as:
\begin{eqnarray}
\label{The planes4}
\alpha^{'}_{i,c} = \arcsin\frac{k_{PIM}}{k_{1,DNMM}}.
\end{eqnarray}
\section{Fresnel formulae in case of absorbing DNMM }
Considering the Fresnel formulae for our absorbing DNMM, one may attempt to proceed as  
Born and Wolf  \cite{born2005} for the metals and  simply use a complex-valued  $\mu$.

Special attention should be given to  the boundary conditions at interfaces, however \cite{valagiannopoulos2015, valagiannopoulos2017}. The point is that in the derivation of Fresnel formulae an important role is played by boundary conditions, which are different for the interface between dielectric media and for the interface between dielectric and conducting (and hence, lossy) media. Namely, 
for an interface between dielectric media, the tangential components of the magnetic vector is continuous, while for dielectric-metallic (or other conducting material) interface the tangential components of the magnetic vector is \textit{discontinuous} and the discontinuity is proportional to the \textit{current surface density}. 
It is thus important to figure out when this discontinuity can be neglected, so that  the Born and Wolf's approach to Fresnel formulae  of absorbing materials can be exploited. Stratton \cite{stratton1941} pointed out  that  the discontinuity is relevant for perfect conductors only, otherwise  to a good approximation the tangential components can be regarded as continuous. 
Hence, we need to estimate  the conductivity of our composite.
There are several  approaches to  describe the effective macroscopic characteristics ( conductivity, permittivity, etc.) of  composite media such as the Maxwell-Garnett theory  (Clausius-Mosotti approximation) \cite{levy1997,garnett1904,landauer1978} and the Bruggeman approximation (the effective medium theory) already mentioned above\cite{bruggeman1935}. We employ the last one  to calculate the effective conductivity of the composite medium, because in this approach all components of the composite are treated on equal footing. The effective conductivity derives as  the root of the following cubic algebraic
equation  (cf. \cite{cai2010}):
\begin{equation}
\label{cubic algebraic equation1}
f_1\frac{\sigma_1 - \sigma_{\eff}}{\sigma_1 + 2\sigma_{\eff}}
+ f_2\frac{\sigma_2 - \sigma_{\eff}}{\sigma_2 + 2\sigma_{\eff}}
+ f_3\frac{\sigma_3 - \sigma_{\eff}}{\sigma_3 + 2\sigma_{\eff}} = 0,
\end{equation}
where $\sigma_1,\sigma_2,\sigma_3, f_1, f_2, f_3$ are the conductivities and the relative concentrations
 of the ingredients 1,2,3 in the composite, respectively, and $\sigma_{\eff}$  is the effective conductivity of the composite.  Generally, the  equation roots depend on the relative concentrations of ingredients.
 For equal concentrations instead of Eq.(\ref{cubic algebraic equation1}) we find the following equation:
\begin{equation}
\label{cubic algebraic equation2}
4\sigma^3_{\eff} - \left(\sigma_1\sigma_2+\sigma_1\sigma_3+\sigma_2\sigma_3\right)\sigma_{\eff} - \sigma_1\sigma_2\sigma_3 = 0.
\end{equation}

For an estimate, suppose that the first ingredient of our mixture is Ag, or Cu  or  Al (cf. \cite{pasko2017}). The conductivities of these metals depend on the EM wave frequency and temperature;  for  room temperatures and our frequency band  one finds the relevant conductivities to be in the same range, namely 
$\sigma_{\mbox{\scriptsize Ag}}= 61.39\times 10^6 \left(\Omega\cdot \mbox{m}\right)^{-1},$
$\sigma_{\mbox{\scriptsize Cu}}=58.6\times 10^6 \left(\Omega\cdot \mbox{m}\right)^{-1},$ and
$\sigma_{\mbox{\scriptsize Al}}=36.69\times 10^6 \left(\Omega\cdot \mbox{m}\right)^{-1}$.
So, assume $\sigma_1=\sigma_{\mbox{\scriptsize Cu}}$.  For iron dioxide (cf. \cite{tralle2014,pasko2017})
and  most of the semiconductors the conductivity is of the order of $10^2 \left(\Omega\cdot \mbox{m}\right)^{-1}$. 
The conductivity of \PbSnTe\ (the third component of our mixture; see \cite{pasko2017}) depends on the Sn-content $x$; here we assume it $10^2 \left(\Omega\cdot m\right)^{-1}$.
Then we have: $p\approx - 2,93\times 10^{9}$, $q\approx - 1.465\times 10^{11}$, $\left(\frac{p}{3}\right)^3 \approx -9.316\times 10^{26}$, $\left(\frac{q}{2}\right)^2\approx 5.365\times 10^{21}$
and hence, $Q < 0$. As a result, three roots of Eq.(\ref{cubic algebraic equation2}) are equal to:
\begin{eqnarray}
\label{three roots}
&& \sigma_{\eff, 1} = 2\sqrt{-\frac{p}{3}} \cos\left(\frac{\phi}{3}\right), \nonumber\\
&& \sigma_{\eff, 2,3} = -2\sqrt{-\frac{p}{3}} \cos\left(\frac{\phi}{3}\pm\frac{\pi}{3}\right),
\end{eqnarray}
where $\cos\phi=-\frac{q}{2\sqrt{-(p/3)^3}}$, $p=-\frac{1}{4}\left(\sigma_1\sigma_2+\sigma_1\sigma_3+\sigma_2\sigma_3\right)$, and $q = -\frac{1}{4}\sigma_1\sigma_2\sigma_3$ 
Thus,  only the first root is positive, while the other two are negative and unphysical.
The numerical value of first root is about $\sigma\approx 5.415\times 10^4 \left(\Omega\cdot m\right)^{-1}$, meaning $\sigma_{\eff} \ll \sigma_1$. 
Our composite is therefore a relatively bad conductor allowing to use the continuity of tangential components
of the magnetic vector as the boundary conditions.
 
For the derivation of Fresnel formulae for the absorptive, magnetic and gyrotropic metamaterial let us consider a slab  of composite metamaterial sandwiched between two layers of dielectric PIM (see Fig.1).
One can  express the reflection and transmission of EM-wave in terms of parameters called reflectivity $\mathcal{R}$ and transmissivity $\mathcal{T}$ via the  coefficients $r_{12}, t_{12}$ and $r_{23}, t_{23}$ associated with the reflection and refraction at the first and second interface respectively.
 Considering at first \textit{TE}-wave (so called \textit{s}-polarization) we find 
  (cf.  \cite{born2005}, \textsection 1.6, (55)-(56))
\begin{eqnarray}
\label{modifying Born and Wolf}
&& r^{\pm}_{12} = \frac{n_1\cos\theta_1 - Z^{-1}_{2(\pm)}\cos\theta_2}{n_1\cos\theta_1 + Z^{-1}_{2(\pm)}\cos\theta_2},
\nonumber\\
&& t^{\pm}_{12} = \frac{2n_1\cos\theta_1}{n_1\cos\theta_1 + Z^{-1}_{2(\pm)}\cos\theta_2}.
\end{eqnarray}
Here $n_1$ is the refraction coefficient of the first PIM and $Z_{2(\pm)} = \sqrt{{\tilde{\mu}^{\pm}_{\eff}}/{\tilde{\epsilon}_{\eff}}}$ is the wave impedance of the metamaterial.
Dropping as previously the subscript $\eff$, we can rewrite the expressions above in an alternative form as
%\begin{widetext}
\begin{eqnarray}
&&r^{\pm}_{12} = \frac{\tilde{\mu}_{2(\pm)}n_1\cos\theta_1 - \tilde{n}_{2(\pm)}\cos\theta_2}{\tilde{\mu}_{2(\pm)}n_1\cos\theta_1 + \tilde{n}_{2(\pm)}\cos\theta_2},\nonumber\\
&&t^{\pm}_{12} = \frac{2\tilde{\mu}_{2(\pm)}n_1\cos\theta_1}{\tilde{\mu}_{2(\pm)}n_1\cos\theta_1 + \tilde{n}_{2(\pm)}\cos\theta_2}.
\end{eqnarray}
%\end{widetext}
%
Here $\tilde{n}_{2(\pm)},\tilde{\mu}_{2(\pm)}$
are the complex refraction index and magnetic permeability of the metamaterial; the indices $+$ and $-$ refer to two values of them (remember, the material is birefringent) while the subscripts $1$  and $2$ refers to the order in which the media are set in this multilayer 'sandwich'. Following \cite{born2005}, we introduce the notation $\tilde{n}_{2(\pm)}\cos\theta_2= u_{2(\pm)} + iv_{2(\pm)}$, where however, $u_{2(\pm)}$ and $v_{2(\pm)}$ have a different form (see below). For the reflection coefficient at the first interface we obtain

\begin{strip}

%\begin{widetext}
\begin{eqnarray}
\label{first interface}
r^{(\pm)}_{12}&=&
\frac{
\left(\mu^{'}_{2(\pm} + i\mu^{''}_{2(\pm)}\right)n_1\cos\theta_1 - \left(u_{2(\pm)}
+ iv_{2(\pm)}\right)}{\left(\mu^{'}_{2(\pm} + i\mu^{''}_{2(\pm)}\right)n_1\cos\theta_1
+ \left(u_{2(\pm)} + iv_{2(\pm)}\right)
}\nonumber\\
&&= \rho^{(\pm)}_{12}\exp\left(i\varphi^{(\pm)_{12}}\right),\nonumber\\
\rho^{(\pm)}_{12}&=&|r^{(\pm)}_{12}|.
\end{eqnarray}
%\end{widetext}
After cumbersome but straightforward calculations we arrive at
%\begin{widetext}	
\begin{eqnarray}
\label{cumbersome1}
\left(
	\rho^{(\pm)}_{12}
\right)^2
&=& \left[
	\frac{
		\left[(\mu^{'}_{2(\pm)})^2 + (\mu^{''}_{2(\pm)})^2\right]
			n^2_1 \cos^2\theta_1
		- \left(
			u^2_{2(\pm)} + v^2_{2(\pm)}
		  \right)
	}{
	(\mu^{'}_{2(\pm)}
	       n_1\cos\theta_1 + u_{2(\pm)})^2
		   + (v_{2(\pm)}+ \mu^{''}_{2(\pm)}
		   )^2
	}
	\right]^2\nonumber\\
&+&
	\frac{
		4n^2_1\cos^2\theta_1(\mu^{'}_{2(\pm)}v_{2(\pm)} - \mu^{''}_{2(\pm)}u_{2(\pm)})^2
	}{
		\left[
			(\mu^{'}_{2(\pm)}n_1\cos\theta_1 + u_{2(\pm)})^2 + (v_{2(\pm)}+ \mu^{''}_{2(\pm)})^2
		\right]^2
	},
\end{eqnarray}
%\end{widetext}
%\begin{widetext}
\begin{eqnarray}
\label{cumbersome2}
\tan\varphi^{(\pm)}_{12} &=&
	\frac{
		2n_1\cos\theta_1\left(\mu^{'}_{}v_{2(\pm)} - \mu^{''}_{2(\pm)}u_{2(\pm)} \right)
	}{
		\left[
			\left(
				\mu^{'}_{2(\pm)}
			\right)^2
			+ \left(
				\mu^{''}_{2(\pm)}
			  \right)^2
		\right]
		n^2_1 \cos^2\theta_1 -
			\left(
				u^2_{2(\pm)} + v^2_{(\pm)}
			\right)
	},
\end{eqnarray}
%\end{widetext}
%\begin{widetext}
\begin{eqnarray}
\label{cumbersome3}
2u^2_{2(\pm)} =
	\left(
		\Re[\tilde{n}_{2(\pm)}]
	\right)^2
	\left(
		1-\delta^2_{\pm}
	\right) - n^2_1\sin^2\theta_1 +
	\sqrt{
		\left[
			\left(\Re[\tilde{n}_{2(\pm)}]\right)^2
			\left(1-\delta^2_{\pm}\right)
			- n^2_1\sin^2\theta_1
		\right]^2
		+4\left(
			\Re[\tilde{n}_{2(\pm)}]
		\right)^4 \delta^2_{\pm}
	},
\end{eqnarray}
%\end{widetext}
%\begin{widetext}
\begin{eqnarray}
\label{cumbersome4}
2v^2_{2(\pm)} &=&
	-\left[
		\left(\Re[\tilde{n}_{2(\pm)}]\right)^2\left(1-\delta^2_{\pm}\right) - n^2_1\sin^2\theta_1
	\right]
	+ \sqrt{
		\left[\left(\Re[\tilde{n}_{2(\pm)}]\right)^2\left(1-\delta^2_{\pm}\right)
		- n^2_1\sin^2\theta_1\right]^2
		+ 4\left(\Re[\tilde{n}_{2(\pm)}]\right)^4\delta^2_{\pm}
	},
\nonumber\\
\delta_{\pm} &=& \frac{n^{''}_{2(\pm)}}{n^{'}_{2(\pm)}}.
\end{eqnarray}
%\end{widetext}
For the transmission at the first surface we obtain:
%\begin{widetext}
\begin{eqnarray}
\label{cumbersome5}
t^{(\pm)}_{12} = |\tau^{(\pm)}_{12}|\exp\left(i\chi^{(\pm)}_{12}\right), \tau^{(\pm)}_{12} = |t^{(\pm)}_{12}|,
\end{eqnarray}
\begin{eqnarray}
\label{cumbersome6}
|t^{(\pm)}_{12}|^2 &= &
	\left[
		\frac{
			\left[\left(\mu^{'}_{2(\pm)}\right)^2 + \left(\mu^{''}_{2(\pm)}\right)^2\right]
			2n^2_1\cos\theta_1
			+ 2n_1\cos\theta_1\left(\mu^{'}_{2(\pm)}u_{2(\pm)} +
			\mu^{''}_{2(\pm)}v_{2(\pm)}\right)
		}{
		\left(\mu^{'}_{2(\pm)}n_1\cos\theta_1 + u_{2(\pm)}\right)^2 + \left(v_{2(\pm)} + \mu^{''}_{2(\pm)}n_1\cos\theta_1\right)^2}
	\right]^2
	\nonumber\\
	&&+ \frac{
		4n^2_1\cos\theta_1\left(\mu^{''}_{2(\pm)}u_{2(\pm)} - \mu^{'}_{2(\pm)}v_{2(\pm)}\right)^2
		}{
			\left[
				\left(\mu^{'}_{2(\pm)}n_1\cos\theta_1 + u_{2(\pm)}\right)^2 + \left(v_{2(\pm)} + \mu^{''}_{2(\pm)}n_1\cos\theta_1\right)^2\right]^2
		}
\end{eqnarray}
%\end{widetext}
%\begin{widetext}
\begin{eqnarray}
\label{cumbersome7}
\tan\chi^{\pm}_{12} =
	\frac{
		2n_1\cos\theta_1\left(\mu^{''}_{2(\pm)}u_{2(\pm)}
		- \mu^{'}_{2(\pm)}v_{2(\pm)}\right)
	}{
		\left[
			\left(\mu^{'}_{2(\pm)}\right)^2
			+ \left(\mu^{''}_{2(\pm)}\right)^2
		\right]
		2n^2_1\cos^2\theta_1
		+ 2n_1\cos\theta_1\left[\mu^{'}_{2(\pm)}u_{2(\pm)} + \mu^{''}_{2(\pm)}v_{2(\pm)}\right]}
\end{eqnarray}
%\end{widetext}
%\end{strip}
For the reflection and refraction coefficients at the first interface  of  \textit{TM}-wave (\textit{p}-polarization) we infer
%\begin{strip}
\begin{eqnarray}
\label{cumbersome8}
r^{(\pm)}_{12}&=&\rho^{(\pm)}_{12}\exp(i\varphi^{(\pm)}_{12}) %\nonumber\\
=\frac{n^{-1}_1\cos\theta_1 - Z_{2(\pm)}\cos\theta_2}{n^{-1}_1\cos\theta_2} %\nonumber\\
=\frac{\tilde{\epsilon}_{2(\pm)}\cos\theta_1-n_1\tilde{n}_{2(\pm)}\cos\theta_2}{\tilde{\epsilon}_{2(\pm)}\cos\theta_1 + n_1\tilde{n}_{2(\pm)}},
\end{eqnarray}
%\end{strip}
\begin{eqnarray}
\label{cumbersome9}
t^{(\pm)}_{12} &=& \tau^{(\pm)}_{12}\exp(i\chi^{\pm}_{12})= \frac{2\cos\theta_1}{\cos\theta_1 + n_1 Z_{2(\pm)}\cos\theta_2}.
\end{eqnarray}
In a similar way  we  derive the explicit expressions for $\rho^{(\pm)}_{12},\varphi^(\pm)_{12}, \tau^{(\pm)}_{12},\chi^{(\pm)}_{12}$ in terms of $\epsilon^{'}_{2(\pm)}, \epsilon^{''}_{2(\pm)}, u_{2(\pm)}, v_{2(\pm)}$ and $\tilde{n}_{2(\pm}$ for the reflection and refraction coefficients for  both waves at the second interface. For example,
%\begin{strip}
\begin{eqnarray}
\label{cumbersome10}
r^{(\pm)}_{23}
	 &=&\frac{
		\left(u_{2(\pm)}-\mu^{'}_{2(\pm)}n_3\cos\theta_3\right)
		+i\left(v_{2(\pm)}-\mu^{''}_{2(\pm)}n_3\cos\theta_3\right)
	}{
		\left(u_{2(\pm)}+\mu^{'}_{2(\pm)}n_3\cos\theta_3\right)
		+i\left(v_{2(\pm)}+\mu^{''}_{2(\pm)}n_3\cos\theta_3\right)} = \rho^{(\pm)}_{23}\exp\left(i\phi^{\pm}_{23}\right), \nonumber\\
\rho^{(\pm)}_{23}&=& |r^{(\pm)}_{23}|.
\end{eqnarray}
%\end{strip}
By means of (\ref{cumbersome10}) one calculates $\tan\phi^{(\pm)}_{23}$. The parameters characterizing the absorbing films (made of DNMM)    which could be measured directly are the reflectivity $\mathcal{R}$, phase shift $\delta_r$ at the reflection, transmissivity $\mathcal{T}$ and phase shift $\delta_t$ on transmission. Using the results obtained above one can derive the corresponding expression for these parameters. It turns out that they are essentially the same as the corresponding formulae of \cite{born2005} but the particular entries are different and determined by the formulae above. For the reflectivity we have
%\begin{strip}
%\begin{widetext}
\begin{eqnarray}
\label{finalR}
\mathcal{R} =
	|r|^2 =
	\frac{
		\left(\rho^{\pm}_{12}\right)^2 e^{2v_{2(\pm)}\eta} + \left(\rho^{\pm}_{23}\right)^2 e^{-2v_{2(\pm)}\eta} +2\rho^{\pm}_{12}\rho^{\pm}_{23}\cos\left(\phi^{\pm}_{23}-\phi^{\pm}_{12} + 2u_{2(\pm)}\eta\right)
	}{
		e^{2v_{2(\pm)}\eta} + \left(\rho^{\pm}_{12}\right)^2\left(\rho^{\pm}_{23}\right)^2e^{-2v_{2(\pm)}\eta} +2\rho^{\pm}_{12}\rho^{\pm}_{23}\cos\left(\phi^{\pm}_{12}+\phi^{\pm}_{23} + 2u_{2(\pm)}\eta\right)
	},\\
\tan\delta_r =
			\frac{\rho^{\pm}_{23}\left(1-\left(\rho^{\pm}_{12}\right)^2\right)\sin\left(2u_{2(\pm)}\eta+\phi^{\pm}_{23}\right)+\rho^{\pm}_{12}\left(e^{2v_{2(\pm)\eta}}-\left(\rho^{\pm}_{23}\right)^2e^{-2v_{(\pm)}\eta}\right)\sin\phi^{\pm}_{12}
		}{
			\rho^{\pm}_{23}\left(1+\left(\rho^{\pm}_{12}\right)^2\right)\cos\left(2u_{2(\pm)}\eta +\phi^{\pm}_{23}\right)+\rho^{\pm}_{12}\left(e^{2v_{2(\pm)\eta}}+\left(\rho^{\pm}_{23}\right)^2e^{-2v_{(\pm)}\eta}\right)\cos\phi^{\pm}_{12}
		}.
\end{eqnarray}
%\end{widetext}
%\end{strip}
Here $\eta=\frac{2\pi}{\lambda_0}h$ where $h$ is the thickness of the film (or slab). These formulae are valid for both waves, $TE$ as well as $TM$; one should simply use for $\rho_{ij}, \phi_{ij}$ the corresponding expressions for $TE$ and $TM$ waves obtained above.
Similarly one can obtain the  expression for the transmissivity $\mathcal{T}$ and the phase shift $\delta_t$ on transmission as
%\begin{strip}
%\begin{widetext}
\begin{eqnarray}
\label{finalT}
&&
\mathcal{T} = \frac{n_3\cos\theta_3}{n_1\cos\theta_1}\frac{\left(\tau^{\pm}_{12}\right)^2\left(\tau^{\pm}_{23}\right)^2\exp\left(-2v_{2(\pm)}\eta\right)}{1 + \left(\rho^(\pm)_{12}\right)^2\left(\rho^(\pm)_{23}\right)^2 e^{-4v_{2(\pm)}\eta} + 2\rho^{\pm}_{12}\rho^{\pm}_{23}e^{-2v_{2(\pm)}\eta}\cos\left(\phi^{\pm}_{12}+\phi^{\pm}_{23}+2u_{2(\pm)}\eta\right)},
\\
&&
\tan\left[\delta_t - \chi^{\pm}_{12} - \chi^{\pm}_{23}+ u_{2(\pm)}\eta\right]
=
\frac{e^{2v_{2(\pm)}\eta}\sin\left(2u_{2(\pm)}\eta\right) - \rho^{\pm}_{12}\rho^{\pm}_{23}\sin\left(\phi^{\pm}_{12}+\phi^{\pm}_{23}\right)
}{
e^{2v_{2(\pm)}\eta}\cos\left(2u_{2(\pm)}\eta\right) + \rho^{\pm}_{12}\rho^{\pm}_{23}\cos\left(\phi^{\pm}_{12}+\phi^{\pm}_{23}\right)
}.
\end{eqnarray}.
%\end{widetext}
\end{strip}
For a $TM$ wave the factor $(n_3\cos\theta_3)/(n_1\cos\theta_1)$ must be replaced by $\left(\cos\theta_3/n_3\right)/\left(\cos\theta_1/n_1\right)$ and for the entries in the last formula
 one should use the corresponding expression for $\tau^{\pm}_{ij}$ and $\chi^{\pm}_{ij}$. Note also that the information concerning refraction index, reflectivity and  transmissivity of absorbing media can be useful for studying the  multiple reflections and
transmissions in a bi-axial slab sandwiched between two anisotropic media \cite{abdulhalim1999}. 
\begin{figure}[h]
\label{fig1}
\includegraphics[width=0.95\linewidth]{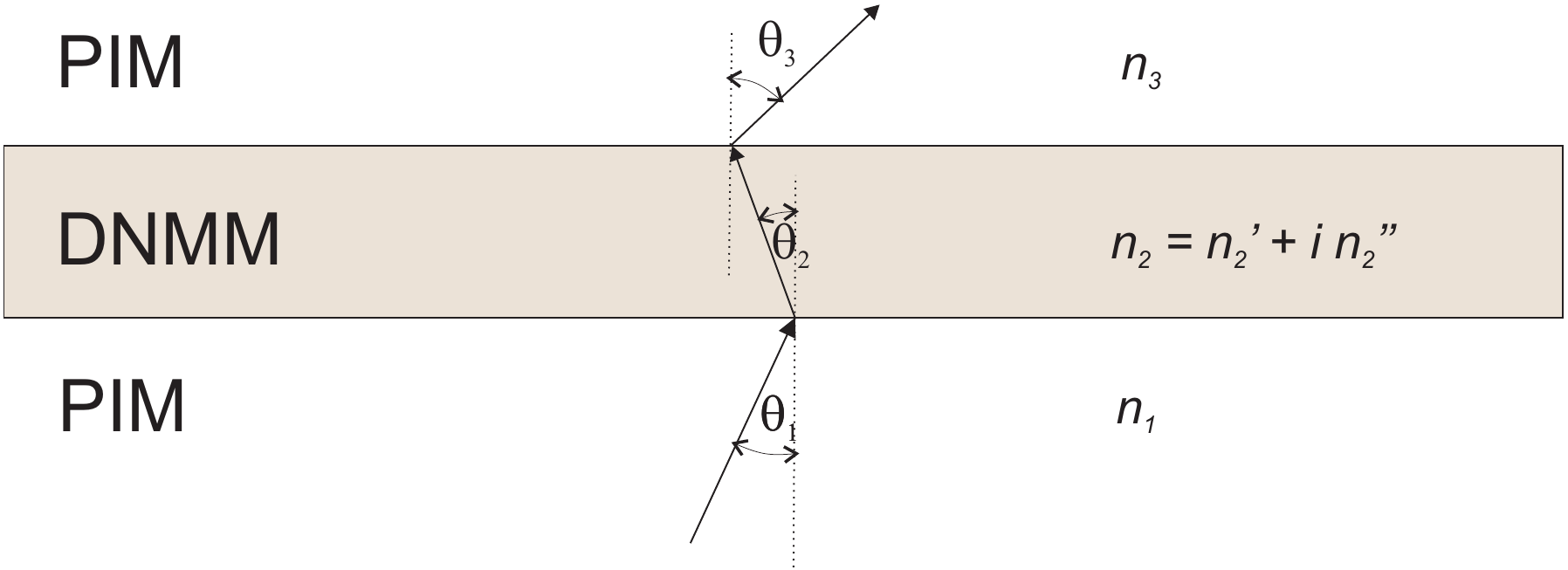}
\caption{Propagation of an electromagnetic wave through a metamaterial film.}
\end{figure}

\begin{figure}[h]
\label{fig2}
\includegraphics[width=0.95\linewidth]{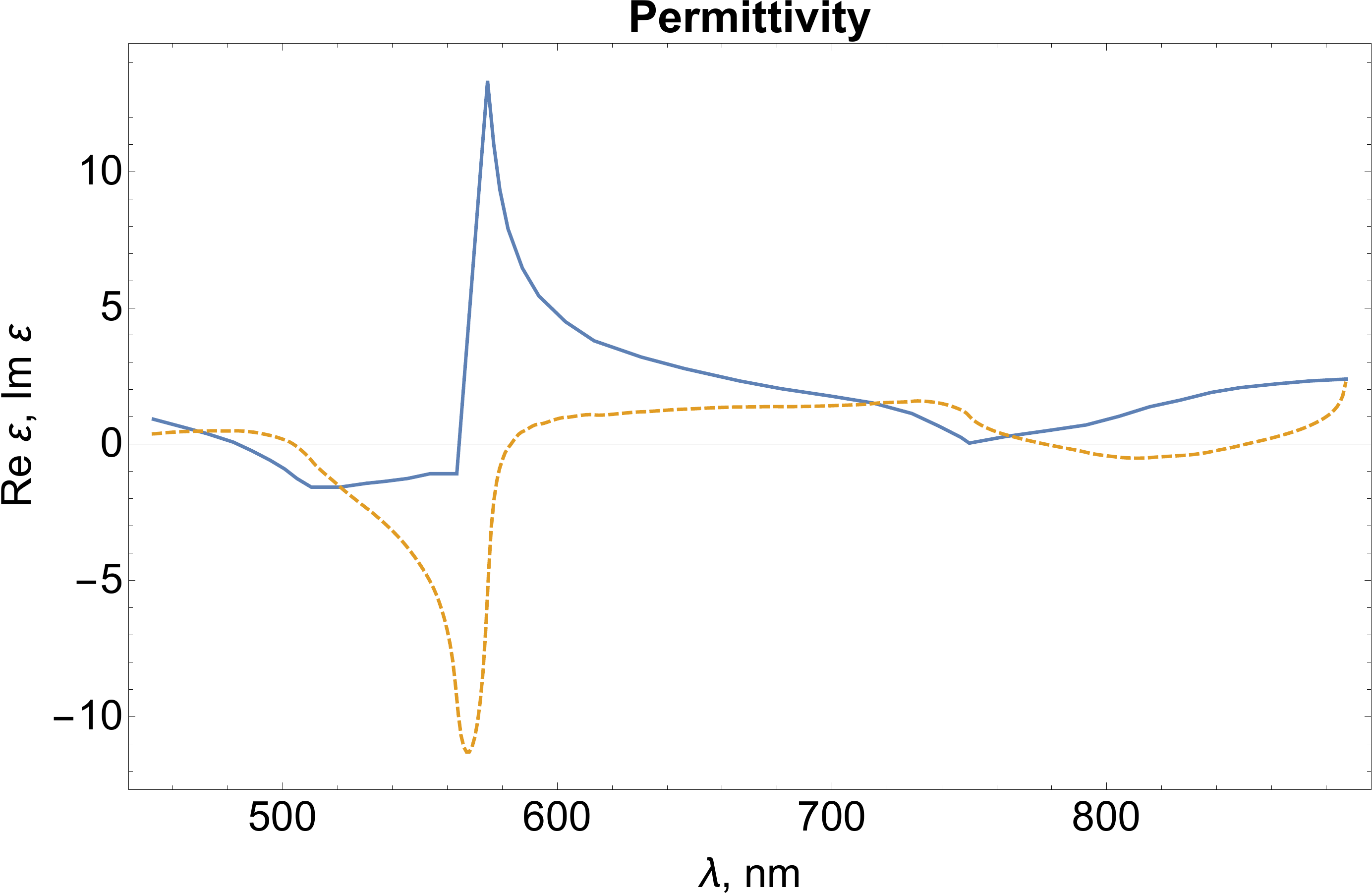}
\includegraphics[width=0.95\linewidth]{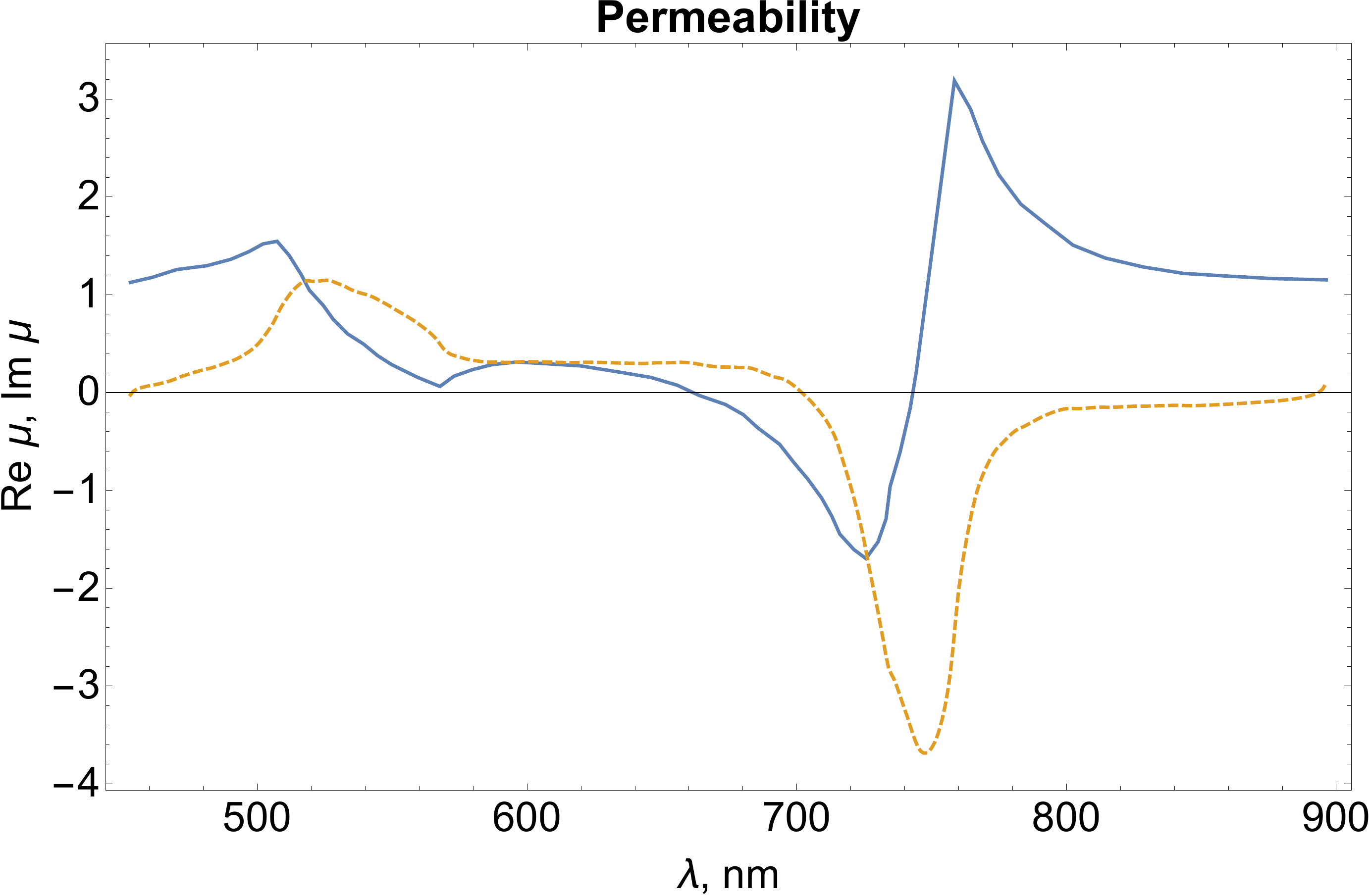}
\caption{Real (solid lines) and reconstructed imaginary parts (dashed lines) of the permittivity and permeability for the sample B of \cite{chattiar2008}.}
\end{figure}

\section{Reflectivity and transmittance of the DNMM-films}

Fig.1 represents schematically the oblique incidence of $EM$-wave on the surface of DNMM sandwiched between the positive refraction index materials.  To demonstrate how these formulae work, one should express the reflectivity and $\mathcal{R}$ transmittance $\mathcal{T}$ in terms of real and imaginary parts of $\epsilon, \mu$ and $n$. It should be noted, that the formulae derived above are valid not only for the metamaterials proposed in \cite{tralle2014,pasko2017}, but for every absorbtive metamaterials which are at the same time relatively bad conductors. The last condition has to be fulfilled for the boundary condition used above at the derivation of Fresnel formulae to be valid. So, in order to compare  the results presented above with experiment, corresponding experimental data are needed. Unfortunately, for various reasons (see \cite{cai2010}, \textsection 4.3 and the references cited therein) these data are scarce; nevertheless, partially some of them can be found \cite{chattiar2008}; they are restricted however only to the real parts of $\epsilon$ and $\mu$.

\begin{figure}[!ht]
\label{fig3}
\includegraphics[width=0.95\linewidth]{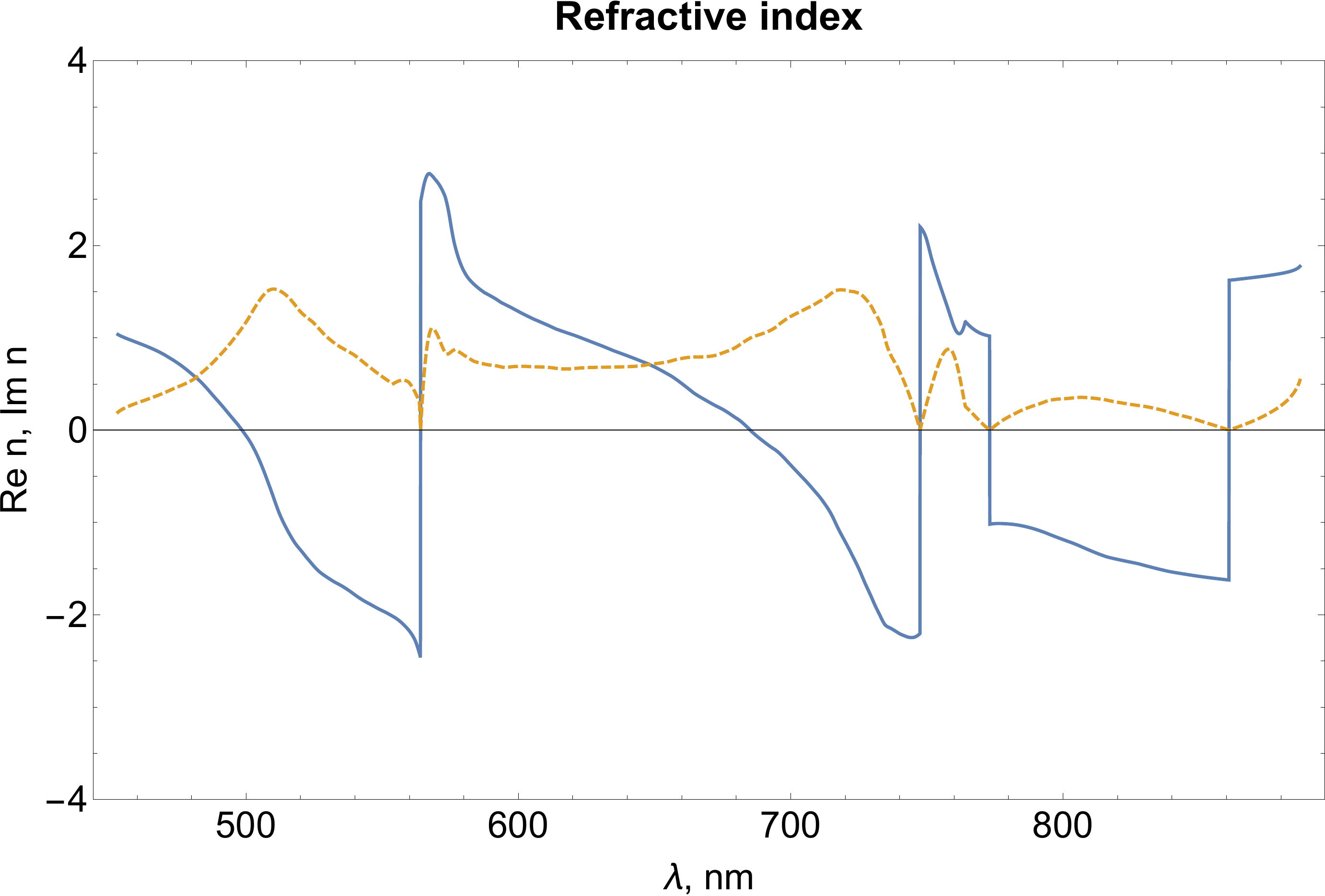}
\caption{Real (solid line) and imaginary part (dashed line) of the refractive index for the sample B of \cite{chattiar2008}.}
\end{figure}
\begin{figure}[!ht]
\label{fig4}
\includegraphics[width=0.95\linewidth]{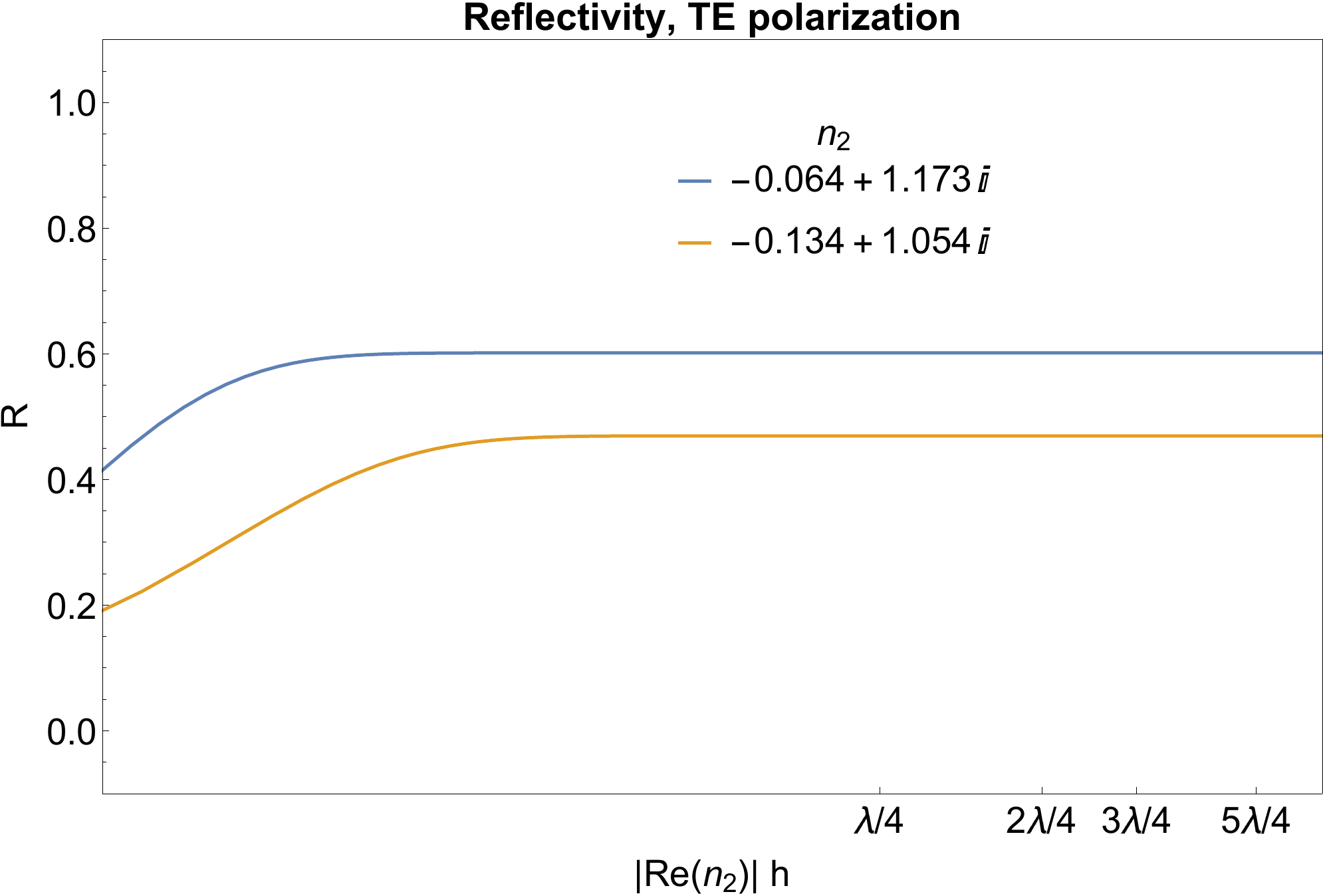}
\caption{Reflectivity of a metamaterial film as a function of its optical thickness for TE polarized wave.}
\end{figure}
\begin{figure}[!ht]
\label{fig5}
\includegraphics[width=0.95\linewidth]{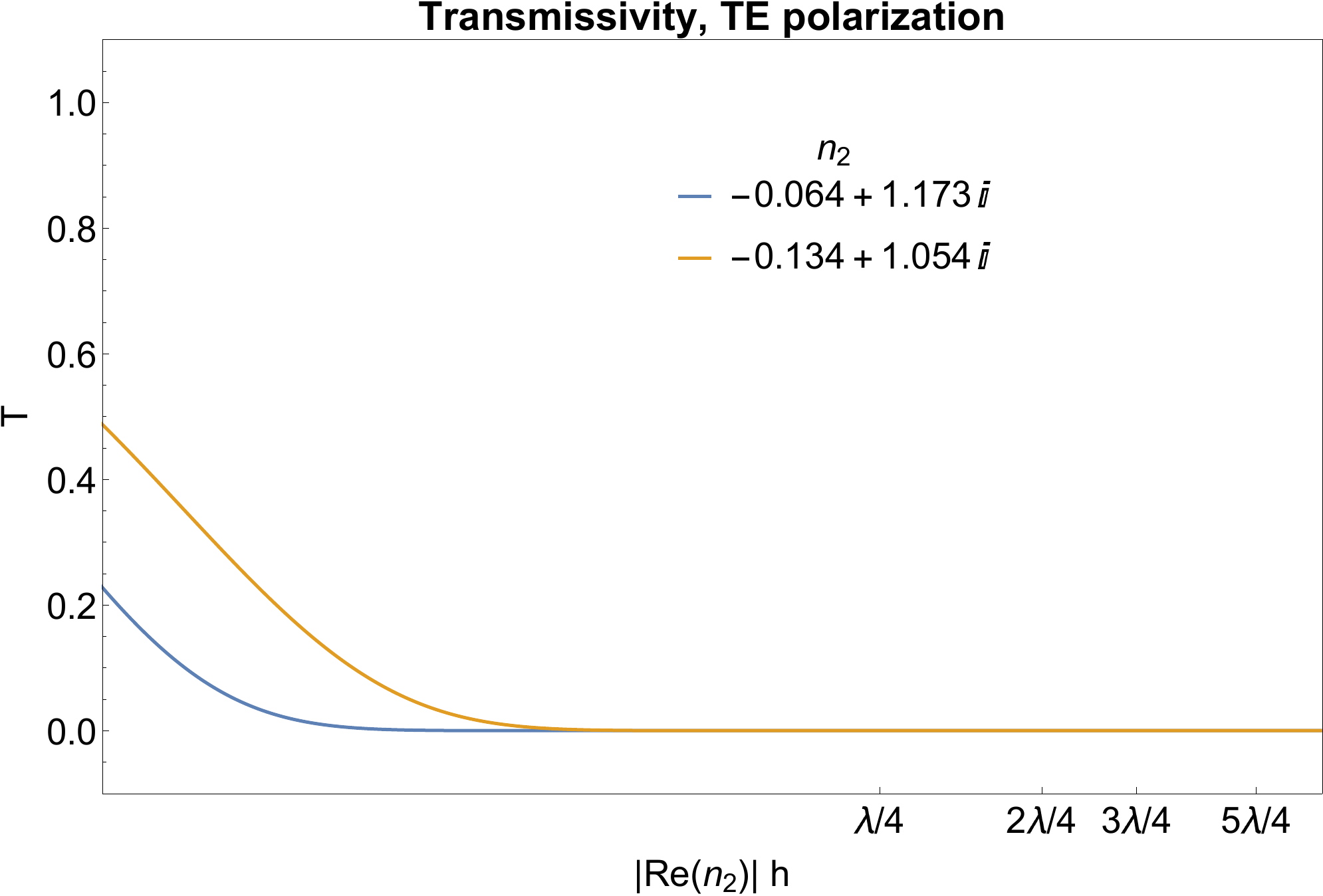}
\caption{Transmissivity of a metamaterial film as a function of its optical thickness for TE polarized wave.}
\end{figure}
\begin{figure}[!ht]
\label{fig6}
\includegraphics[width=0.95\linewidth]{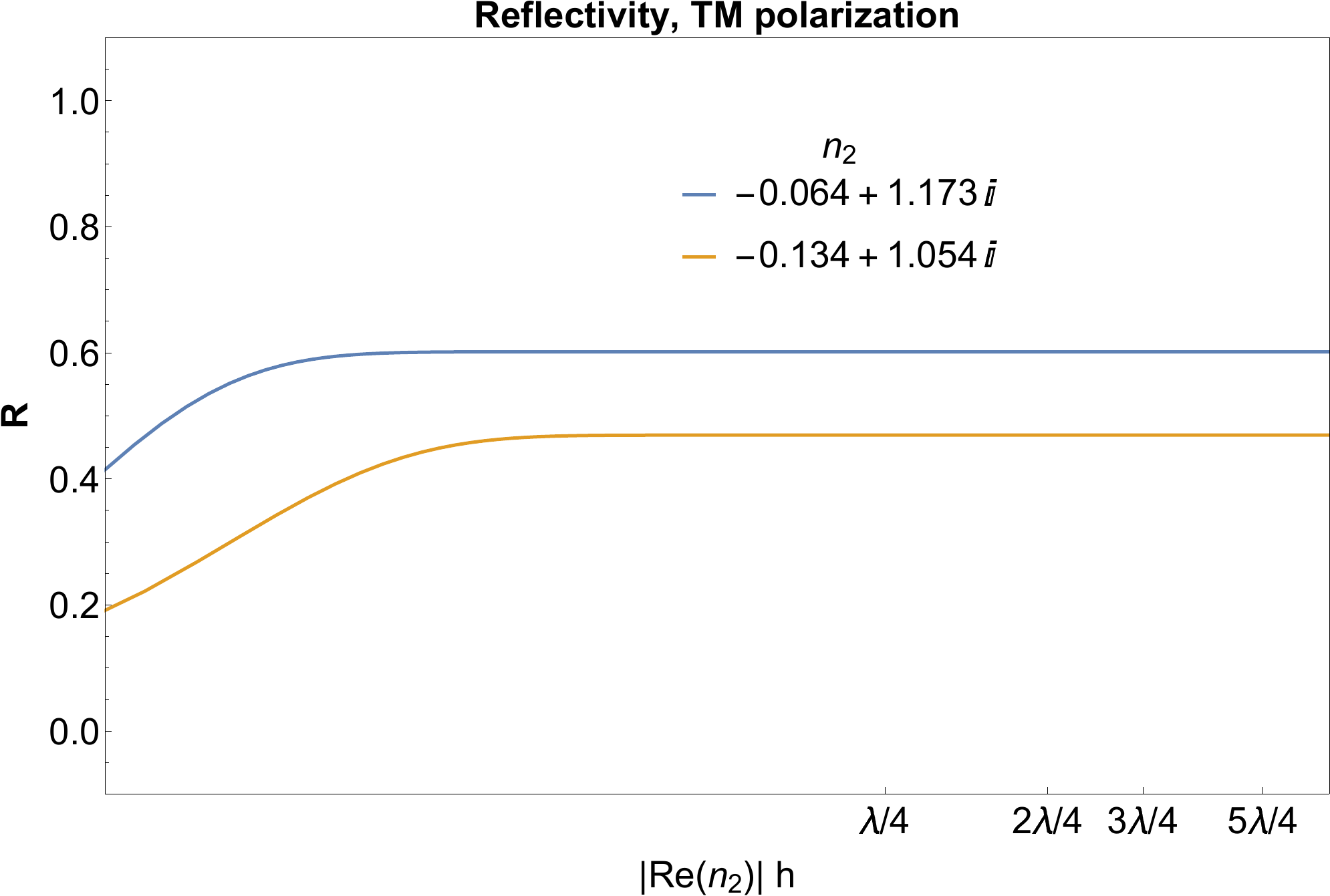}
\caption{Reflectivity of a metamaterial film as a function of its optical thickness for TM polarized wave.}
\end{figure}
\begin{figure}[!ht]
\label{fig7}
\includegraphics[width=0.95\linewidth]{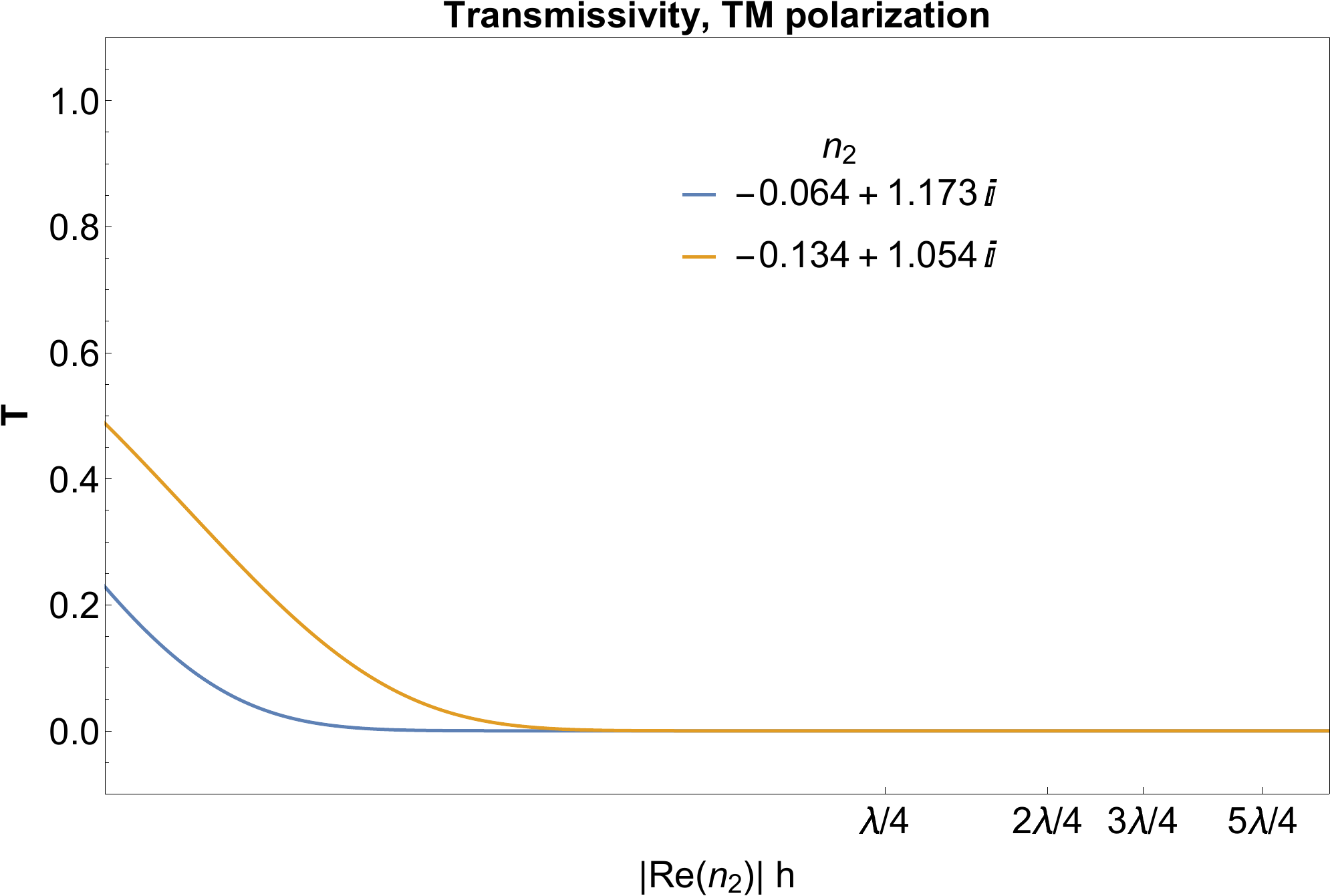}
\caption{Transmissivity of a metamaterial film as a function of its optical thickness for TM polarized wave.}
\end{figure}

\begin{figure}[!ht]
\label{fig8}
\includegraphics[width=0.9\linewidth]{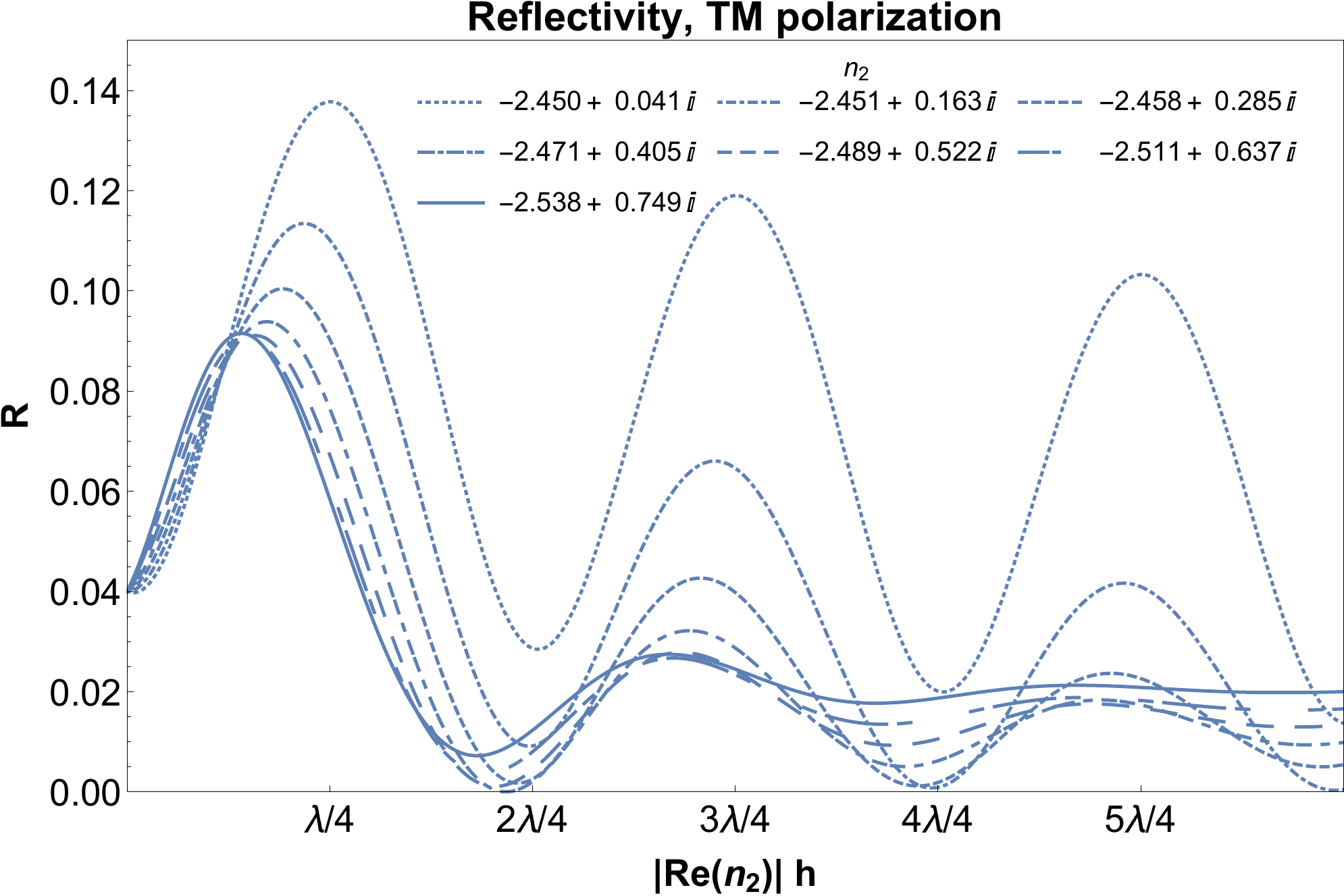}
\caption{Reflectivity of a metamaterial film as a function of its optical thickness for TE polarized wave.}
\end{figure}
\begin{figure}[!ht]
\label{fig9}
\includegraphics[width=0.9\linewidth]{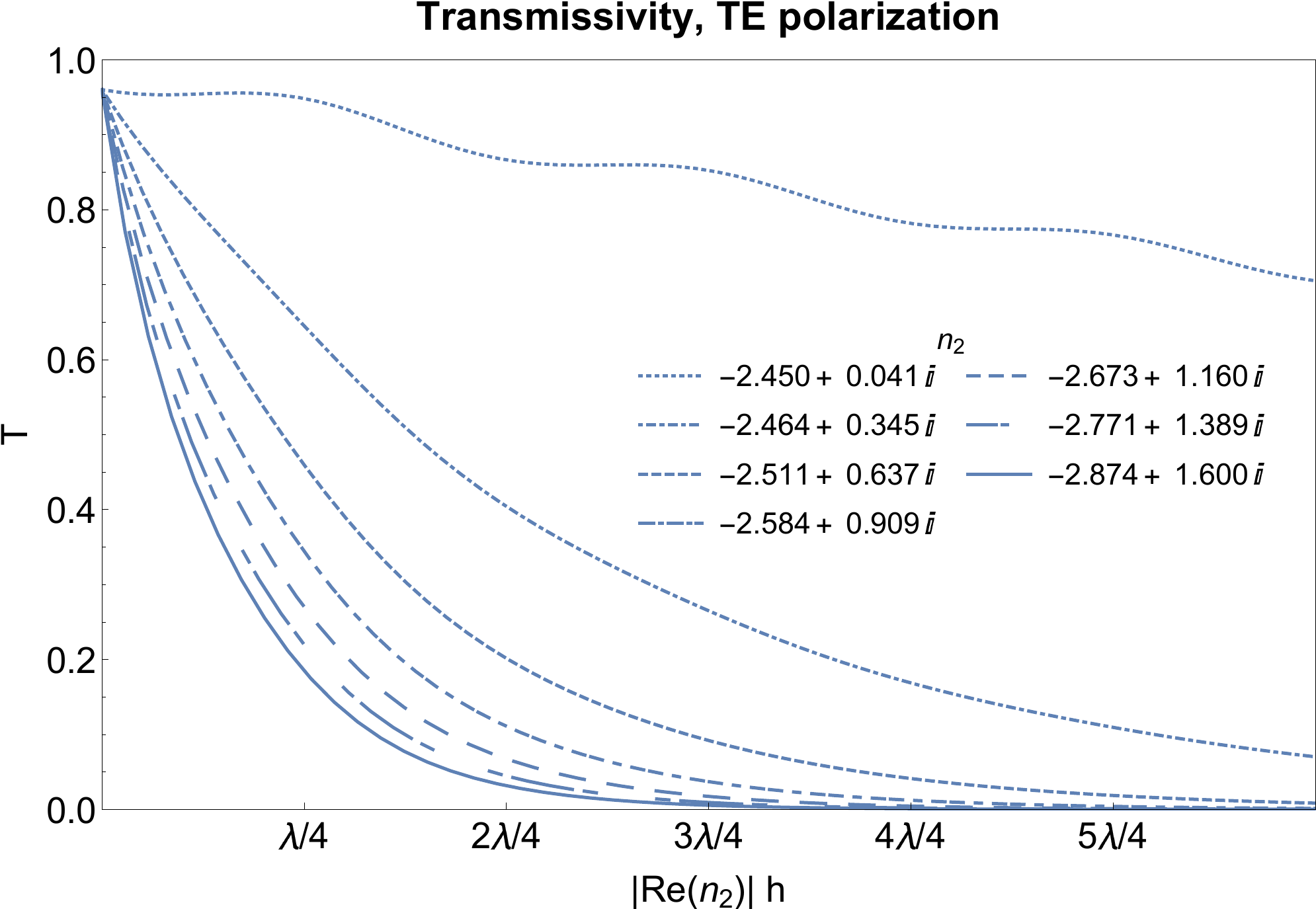}
\caption{Transmissivity of a metamaterial film as a function of its optical thickness for TE polarized wave.}
\end{figure}

\begin{figure}[!ht]
\label{fig10}
\includegraphics[width=0.9\linewidth]{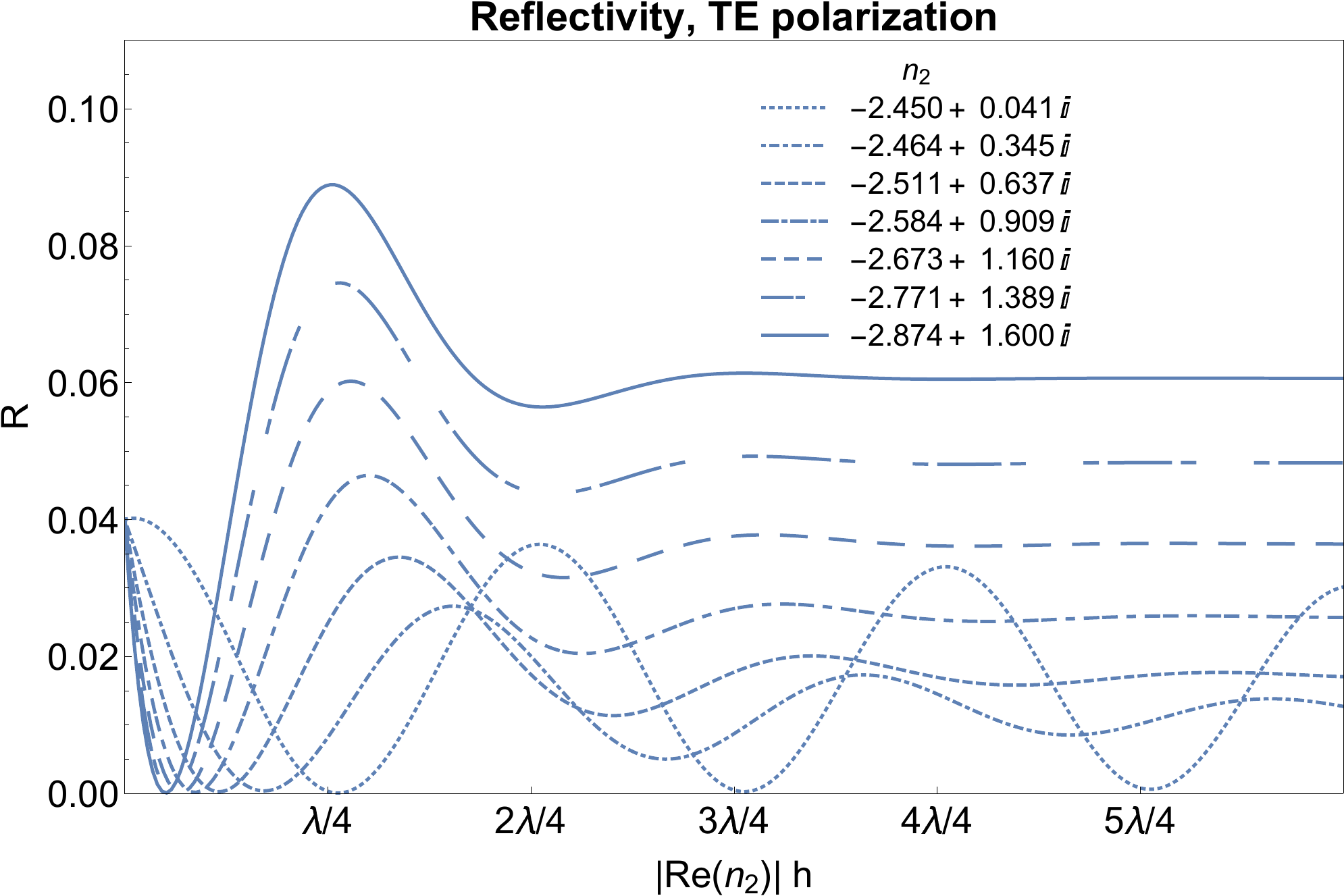}
\caption{Reflectivity of a metamaterial film as a function of its optical thickness for TM polarized wave.}
\end{figure}
\begin{figure}[!ht]
\label{fig11}
\includegraphics[width=0.9\linewidth]{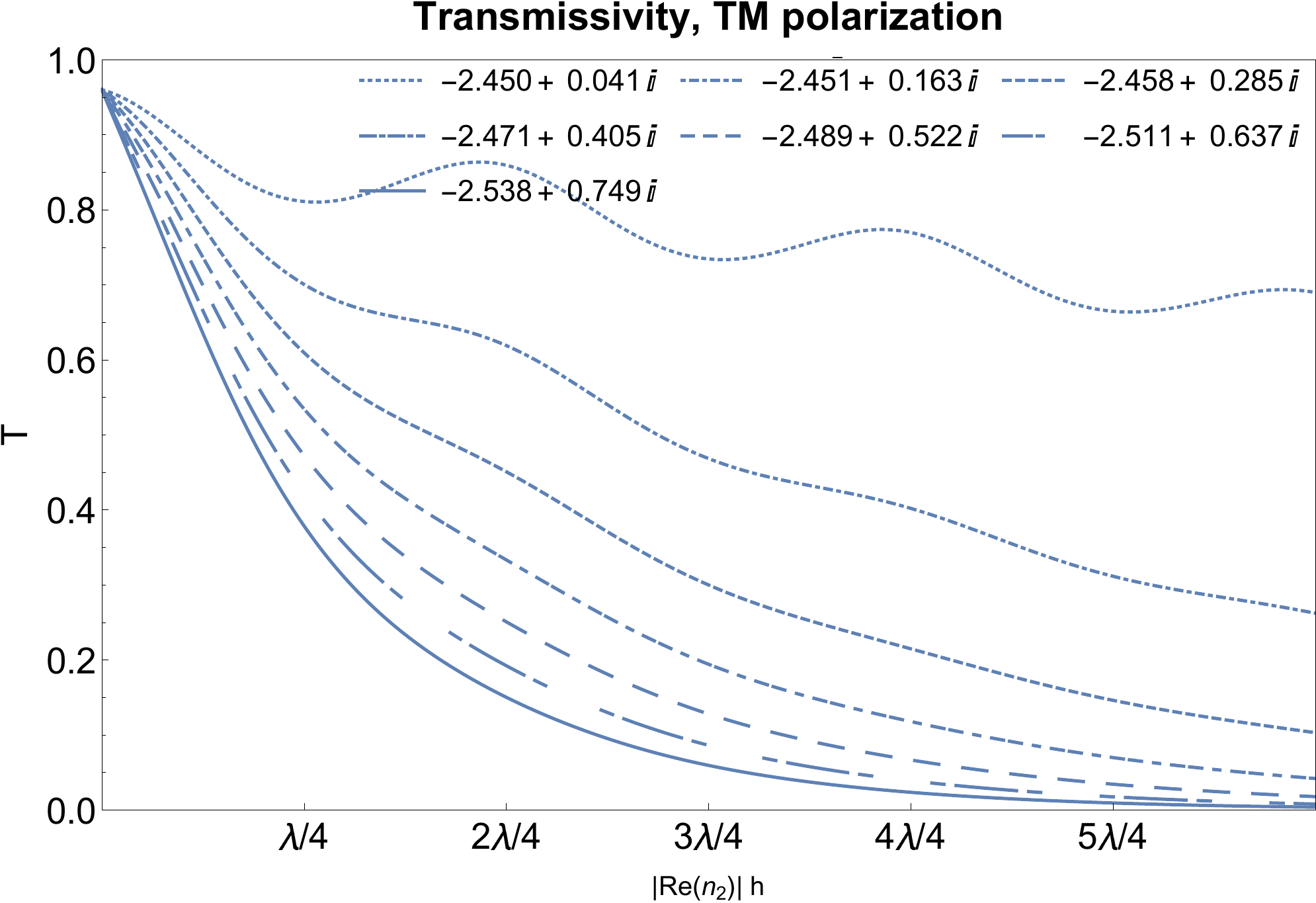}
\caption{Transmissivity of a metamaterial film as a function of its optical thickness for TM polarized wave.}
\end{figure}

Using Kramers-Kronig relations % (which is nothing else but the special case of Hilbert integral transform):
\begin{eqnarray} 
&&
\chi_1(\omega) = \frac{1}{\pi} \it{VP} \int_{-\infty}^{\infty}\frac{\chi_2(\omega^{'})}{\omega^{'} - \omega}d\omega^{'},\\
&&
\chi_2(\omega) = \frac{1}{\pi} \it{VP} \int_{-\infty}^{\infty}\frac{\chi_1(\omega^{'})}{\omega^{'} - \omega}d\omega^{'},
\end{eqnarray}
where $\chi(\omega) = \chi_1(\omega) + i\chi_2(\omega)$ is complex-valued function, while $\chi_1(\omega)$, $\chi_2(\omega)$ both are real and \textit{VP} stands for Cauchy principal value. Then, we were able to calculate the real and imaginary parts of the refraction index for broad frequency domain. The values of real parts of $\epsilon$ and $\mu$ were taken from the work \cite{chattiar2008}.
The results of calculations are presented in Figs.~2 and~3.
To illustrate formulae (\ref{finalR}) and (\ref{finalT}) only two values of $n_2 = - 0.064  + i\times 1.173$ and $n2 = - 0.134  + i\times 1.054$,  each corresponding to different frequency, were chosen to draw the plots in Figs.~4-7. The choice of these two frequencies  were not determined by some reasoning, and this example is only illustrative one; we would like to demonstrate how the explicit  Fresnel formulae work in the particular case of metamaterial. As it easily can be seen, the imaginary part of refraction index is positive everywhere in considered frequency domain, as it of course should be in accordance with causality condition. 
For clarity and readers convenience, we present the results of calculations made for permeability, refraction index, as well as the reflectivity and transmissivity for both, TE and TM polarizations on the separate charts.  These results could be checked directly by proper future experiments.

As for the Figs.~8-11, we were interested  to compare  the  reflectivity and transmissivity of the film made of metamaterial with these values for dielectric films considered by Born and Wolf (\cite {born2005}, page 68, Fig. 1.18)  in order to observe whether  these parameters would periodically dependent on the film thickness or not. To this end, we modeled some fictitious  metamaterial choosing the corresponding values of permittivity and permeability and made the calculations as it was described above. As it can be seen in the Figs.~8-11, in case of available experimental data used in our simulations, for the trasmissivity of both TE and TM waves such quasi-periodic dependence is practically absent or at least very weak, while for the reflectivity  it is clearly seen for some values of real and imaginary parts of complex refraction index and as it is observed for some metal films. Note also, that the values in horizontal axis in Fig. 4-11 are expressed as $Re [n_2]\times h$ where $n_2$ and $h$ are the refraction index and thickness of metamaterial layer sandwiched between the positive refraction index materials as it is shown in Fig.~1. Additionally, they are presented  in log-scale in Figs.~4-7.

\section{Conclusions}

In this paper we studied the optical characteristics of double-negative metamaterials taking into account the fact that all metamaterials are inevitably absorptive and derived the explicit formulae for total reflection angle as well as correct Fresnel formulae describing the reflection and refraction coefficients for the DNMM in case of \textit{TE} as well as \textit{TM} EM wave polarization. The reflectivity and transmittance of DNMM film embedded in the PIM-surrounding  are also presented.

\section{Acknowledgment}

This work has been done due to the support which two of us (IT and PZ) have got from the Centre for Innovation and Transfer of Natural Science and Engineering Knowledge, University of Rzesz\'ow.

\section*{CRediT authorship contribution statement}

\textbf{Igor Tralle}: Conceptualization, Formal analysis, Writing – Original draft preparation, Supervison. 
\textbf{Levan Cho\-tor\-lish\-vi\-li}: Methodolgy,  Writing - Review \& Editing.
\textbf{Paweł Zi\c{e}ba}: Software, Visualization,  Writing - Review \& Editing.

\end{document}